\documentclass[12pt]{article}
\usepackage{amssymb,amsmath,amsfonts}
\usepackage{graphicx}
\graphicspath{{Figures/}}
\usepackage{color}
\usepackage{enumitem}
\numberwithin{equation}{section}
\newcommand{\re}{\mathrm{Re}\, }
\newcommand{\im}{\mathrm{Im}\,}
\newcommand{\I}{{\rm{i}}}
\newcommand{\D}{{\rm{d}}}
\newcommand{\E}{{\rm e}}
\newcommand{\tr}{\operatorname{tr}}

\newcommand{\dss}{\displaystyle}

\newcommand{\be}{\begin{equation}}
\newcommand{\ee}{\end{equation}}
\newcommand{\bea}{\begin{eqnarray}}
\newcommand{\eea}{\end{eqnarray}}

\newcommand{\supp}{\operatorname{supp}}

\newcommand{\intmult}[1]{\operatornamewithlimits{\idotsint}_{#1}}

\newcommand{\ket}[1]{| #1 \rangle}
\newcommand{\bra}[1]{\langle #1 |}

\newcommand{\ketbra}[2]{| #1 \rangle \langle #2 |}
\newcommand{\scalprod}[2]{\left\langle {#1}, {#2}\right\rangle }

\def\real{{\mathbb{R}}}

\def\complex{{\mathbb{C}}}

\def\proba{{\rm I\kern -.18em P}}

\newcommand{\idop}{\mathchoice {\rm 1\mskip-4mu l} {\rm 1\mskip-4mu l}
	{\rm 1\mskip-4.5mu l} {\rm 1\mskip-5mu l}}
\newcommand{\eps}{\varepsilon}
\newcommand{\Proof}{\noindent {\bf Proof. }}
\newcommand{\Proofof}[1]{\noindent {\bf Proof of #1. }}
\newcommand{\finpro}{\hfill $\Box$}

\newcommand{\ie}{i.e.\;}

\newcommand{\RHS}{right-hand side\;\,}
\newcommand{\LHS}{left-hand side\;\,}
\newcommand{\onehalf}{\frac{1}{2}}

\newcommand{\vv}{{\underline{v}}}

\newcommand{\vvv}{{\underline{\mathbf{v}}}}

\newcommand{\Dd}{{\cal D}}

\newcommand{\Oo}{{\cal O}}

\newcommand{\HS}{H_S}
\newcommand{\HR}{H_R}
\newcommand{\Hint}{H_{\rm int}}
\newcommand{\U}{U_{\lambda,\varepsilon}}
\newcommand{\UA}{V_{\lambda,\varepsilon}}
\newcommand{\WK}{W_{\rm K}}
\newcommand{\UPsi}{\Psi_{\lambda,\varepsilon}}
\newcommand{\X}[1]{X_{#1}}
\newcommand{\Waveop}{\Omega_{\lambda,\varepsilon}}
\newcommand{\Kt}[2]{\widetilde{K}_{#1 #2}}
\newcommand{\dynphase}[2]{\varphi_{#1 #2}}
\newcommand{\transproba}{p_{1 \rightarrow 2}^{(\lambda,\varepsilon)}}
\newcommand{\transprobafree}{p_{1 \rightarrow 2}^{(0,\varepsilon)}}
\newcommand{\qtrans}{q_{1 \rightarrow 2}}
\newcommand{\omeg}[1]{\omega^{(#1)}_{\lambda,\eps}}
\newcommand{\tp}{t_{\rm p}}
\newcommand{\Rrj}{{\mathcal R}_{\lambda,\eps}^{(j)}}

\newcounter{resultcounter}[section]

\newtheorem{thm}[resultcounter]{Theorem}
\newtheorem{lem}[resultcounter]{Lemma}
\newtheorem{prop}[resultcounter]{Proposition}
\newtheorem{cor}[resultcounter]{Corollary}

\newtheorem{rmk}[resultcounter]{Remark}

\begin{document}


\title{Adiabatic transitions in a two-level system coupled to a free Boson reservoir}
\author{Alain Joye\footnote{Univ. Grenoble Alpes, CNRS, Institut Fourier, 38000 Grenoble, France; Alain.Joye@univ-grenoble-alpes.fr} , 
  Marco Merkli\footnote{Memorial University of Newfoundland, Department of Mathematics and Statistics, St. John's, NL, Canada A1C 5S7; merkli@mun.ca} , and
  Dominique Spehner\footnote{Universidad de Concepci\'on, Facultad de Ciencias F\'{\i}sicas y Matem\'aticas,
    Departamento de Ingenier\'{\i}a Matem\'atica, Chile {\&}  Univ. Grenoble Alpes, CNRS, Institut Fourier and LPMMC, 38000 Grenoble, France; dspehner@ing-mat.udec.cl}
}

\maketitle
		
\begin{abstract}
 We consider a time-dependent two-level quantum system interacting with  a free Boson reservoir.
The coupling is energy conserving and depends slowly on time, as does the system Hamiltonian, with a common adiabatic parameter $\eps$.
 Assuming that the  system and reservoir are initially decoupled, with the reservoir in equilibrium at temperature $T\ge0$, 
  we compute the transition probability from one eigenstate of the two-level system to the other eigenstate as a function of time, in the regime of small
  $\eps$ and small coupling constant $\lambda$. We analyse the deviation from the adiabatic transition probability obtained in absence of the reservoir.
\end{abstract}

\section{Introduction}

In this paper we study the transition probability between the energy
eigenstates of a driven two-level 
system in contact with an environment, a bosonic reservoir at zero {or at positive}
  temperatures.
The Hamiltonian of the two-level system and the coupling with
  the reservoir both depend on time, 
  varying  on a slow time scale $1/\eps$; that is, they are functions of the rescaled time $t=\eps \tp$, where $\tp$ is the physical time.
  We consider
  interaction Hamiltonians which are linear in the bosonic field operators and for which the system and reservoir do not exchange energy instantaneously, meaning that the system Hamiltonian commutes with the interaction at any given time. 
  
The initial system-reservoir state is taken to be disentangled, with the two-level system in an eigenstate of its Hamiltonian and the reservoir in equilibrium at temperature $T\ge0$. Our main goal is to determine the probability, denoted $\transproba (t)$, to find the
system in the other
eigenstate at  some fixed rescaled time $t>0$. We do this in the adiabatic and weak coupling regime, meaning that  $\eps$ and
the system-reservoir coupling constant $\lambda$ are both small.

The adiabatic regime yields rather detailed and precise approximations of the true quantum dynamics in a variety of physically relevant situations and its study has a long history. The adiabatic theorem of quantum mechanics was first stated for self-adjoint time-dependent Hamiltonians with isolated eigenvalues in \cite{BF, Kato50}, and then extended to accommodate isolated parts of the spectrum, see \cite{N, ASY}. This version applies to the two-level system we consider, in absence of coupling. Adiabatic approximations for gapless Hamiltonians,  where the eigenvalues are not isolated from the rest of the spectrum, were later established in \cite{AE, T}. This is in particular the situation for the total Hamiltonian of the two-level system coupled to a free boson reservoir.
Then, adiabatic theorems were formulated in \cite{A-SF, J2, AFGG1} for dynamics generated by non-self-adjoint operators, leading to extensions of the gapless, non self-adjoint case in \cite{Sch}. Such results apply to the dynamics of open quantum system within the markovian approximation, by means of time dependent Lindblad generators. Finally, the adiabatic approximation was also shown to be exponentially accurate for analytic time dependence \cite{JKP, JP, N2, Joye}, in line with the famous Landau-Zener formula; see \cite{HJ} for more details.

Applied to our two-level system without coupling to the reservoir ($\lambda=0$, isolated eigenvalues), the adiabatic theorem says that the transition probability  $\transprobafree (t)$ is  of order $\eps^2$.
By contrast, the gapless adiabatic theorem applied to the total
Hamiltonian of the system and reservoir
in general tells us merely that the transition probability is $o(\eps)$~\cite{AE,T}. 

We show that in our model,  $\transproba (t)$ 
differs from the transition probability $\transprobafree(t)$ (no coupling)  by a term of order $\eps \lambda^2$, which
we  determine explicitly. At zero temperature, this term turns out to be nonzero when the transition is from the upper to the lower energy level, while it vanishes for the reverse transition, up to corrections of higher orders in  $\eps$ and $\lambda$. At positive temperatures,
 it is nonzero  for both transitions.  We also identify parameter regimes in which this correction term is the leading one of $\transproba (t)$.

To our knowledge, the problem we address here  has  been studied in the  mathematical physics literature only by means of an effective
description of the open system, namely, employing a time-dependent Lindblad operator \cite{AFGG1, Avron10, Fraas16}.
For a dephasing Lindblad operator, commuting with the generator of the system Hamiltonian,
the authors there determine the transition
probability between distinct energy levels of the
system in the adiabatic limit.  Like in our microscopic model, they find that this probability is of  order $\eps$, but in their
 Lindbladian approach, the dependence of the probability on whether it is up- or downwards does not show.

Let us also mention that the general theme addressed here is relevant for the discussion of the validity of the Born-Oppenheimer approximation in presence of a scalar photon field. See for example \cite{TW} which provides a detailed analysis of this type of questions  in a regime where the effect of the field is a lower order correction. In spirit, it corresponds in our setting to the regime $\lambda \ll \sqrt{\eps}$ (see the discussion in Remark \ref{rmk-2}).

\section{Model and main result}
\subsection{The model}

Let us start by describing the model at  zero temperature, see Section \ref{postemp1} for the positive temperature case. 	
To account for its slowly-varying nature, the self-adjoint system Hamiltonian $\HS ({ \eps \tp})\in M_2(\complex)$  at physical time $\tp$ is assumed to be a function of   
the  rescaled time $t= \eps \tp \in [0,1]$,
with $\eps$ a small, positive parameter; $\eps\rightarrow 0$
is  the adiabatic limit.
The Hilbert space of the total system is
 \begin{equation}\label{m1}
 {\mathcal H}_{\rm tot}=\complex^2\otimes \mathcal{F}_{+}(L^2(\real^3)) \; ,
 \end{equation}
 where $ \mathcal{F}_{+}(L^2(\real^3))$ denotes the bosonic Fock space on $L^2(\real^3)$, the Hilbert space in three dimensional momentum space. 
 The coupling to the reservoir is linear in the bosonic field operator
\begin{equation}\label{fieldop}
  \phi (g) = \frac{1}{\sqrt{2}} \int_{\real^3} \D^3 k \big( \overline{g(k)} a (k) + g(k) a^\ast (k) \big) ,
\end{equation} 
where $g \in L^2 (\real^3)$ is the form factor and $a^\ast(k)$, $a(k)$ are the creation and annihilation operators of a boson with momentum $k$. 
The system-reservoir interaction Hamiltonian is 
\begin{equation} \label{H_int}
  \Hint (\eps \tp ) = \lambda B ( \eps \tp ) \otimes \phi (g)\;,
\end{equation}
where $\lambda>0$ is the coupling constant and $B(\eps \tp )$ is a slowly-varying self-adjoint operator on $\complex^2$, varying on the same timescale as the system Hamiltonian.
We assume that $[\HS(t),B(t)] = 0$ for all $t \in [0,1]$.
This means that there are no instantaneous energy exchanges between the system and reservoir.
The two self-adjoint operators $\HS ( t )$ and $B(  t )$ can thus be diagonalized simultaneously, 
\begin{equation} \label{H_S}
\HS (  t  ) = \sum_{j=1}^2 e_j(  t ) P_j(  t ) ,\quad B(  t )=  \sum_{j=1}^2 b_j( t ) P_j( t )\;,
\end{equation}
where $\{ P_j(t)\}_{j=1}^2$ is a complete set of orthogonal projections on $\complex^2$
and $e_j(  t )$, $b_j( t )$ are real eigenvalues depending on the rescaled time $t$.

In what follows, we set
\begin{equation}
  e_{21} (t) = e_2(t)- e_1(t) \; , \qquad b_{21}(t)=b_2(t)-b_1(t)\;.
\end{equation}  

We shall rely on standard assumptions in the context of adiabatic theorems 
on both self-adjoint operators $\HS$ and $B$:

\begin{enumerate}[label=(A.\arabic*)]
\item  \label{assum-gap} {Gap hypothesis:}  $\delta = \inf_{t \in [0,1]} | e_{21} (t)|  >0$.
\item  \label{assum-smoothness}
  The eigenvalues $e_j(t)$ and $b_j(t)$ and 
  spectral projectors $P_j(t)$ are of class $C^4 (]0,1[)$, with all derivatives having well defined limits at $\{0,1\}$.
\item \label{assum-constant_eigenvectors}
  The eigenprojectors satisfy $\lim_{t\rightarrow 0^+}\partial_t^nP_j(t)=0$ for all $n\in\{1,\dots,4\}$.
\end{enumerate}

Thanks to assumption \ref{assum-gap}, the spectral projectors
$P_j(t)$ are rank one at all times, so that $P_j(t)= \ketbra{\psi_j(t)}{\psi_j(t)}$ with
$\{ \psi_j(t) \}_{j=1,2}$ an orthonormal common eigenbasis of $\HS( t )$ and $B( t )$ 
that can be chosen to be $C^4 (]0,1[)$.
When clear from the context, we often write $Z(t)$ instead of $Z(t) \otimes \idop$ for operators $Z(t)$ on
$\complex^2$.

The time-independent Hamiltonian of the bosonic reservoir reads
\begin{equation}\label{m12}
\HR = \int \D^3 k \, \omega (k) a^*(k) a(k) \;.
\end{equation}
We will assume $\omega$ to depend only on the modulus $|k|$ of the wave vector $k$.

The system and bosons are uncorrelated at time $t=0$ and initially in the eigenstate
$\psi_1(0)$ of $\HS (0)$ and in the vacuum state  $\chi \in {\mathcal F}_{+}(L^2(\real^3))$, respectively.  Hence, the initial state of the
system and of the zero-temperature reservoir is
\begin{equation} \label{eq-init_state}
  \rho (0) = \ketbra{\psi_1(0)}{\psi_1(0)} \otimes \ketbra{\chi}{\chi}\;,
\end{equation}
where, for any vectors $\mu, \nu$, $\ketbra{\mu}{\nu}$ denotes the rank one operator $\eta\mapsto  \langle \nu | \eta \rangle \mu$.
The system-reservoir evolution operator $\U (t)$   is given by the time-rescaled Schr\"odinger equation
\begin{equation} \label{eq-Schrodinger_Ueps}
  \I \eps \partial_t \U (t) = \Big( \HS (t) \otimes \idop + \lambda B(t) \otimes \phi (g) + \idop \otimes \HR \Big) \U (t) \; , \ \ \  \U (0 ) = \idop\;, \ \ \ t\in (0,1)\;,
\end{equation}  
the operator inside the brackets being the total Hamiltonian.
Here and below, all derivatives are understood in the strong sense, on  $\Dd=\complex^2\otimes \Dd_R$, where $\Dd_R$ is the domain of $H_R$.
 The reduced state on the two-level system is given by taking the partial trace,
 \begin{equation} 
\rho_S(t)=\tr_R [ U_{\lambda,\eps} (t)\rho(0) U_{\lambda,\eps}^*(t)].
\end{equation}  

We note here that the coupling with the reservoir in our model leads to a pure dephasing type evolution for the reduced state 
$\rho_S(t)$  when $\HS$ and $B$ are time-independent. In this case, for any initial system-reservoir product state $\rho(0)= \rho_S \otimes \rho_R$, if
$(\rho_S(t))_{kj}$ denotes the matrix elements of the reduced state in a common eigenbasis of $\HS$ and $B$, the level populations $(\rho_S(t))_{jj}$ are time independent
while the off-diagonal elements decay with time.
Although there is no energy exchange and no relaxation towards an equilibrium, the coupling with the reservoir induces 
decoherence in the system, so one  says that the system undergoes a pure dephasing dynamics.
The situation is different for driven systems where $\HS(t)$ and $B(t)$ are time dependent and commute at all times, and we seek to quantify the transition probability between instantaneous energy levels.

\subsection{Adiabatic transition probability}

%
The transition probability  of the system from level $1$ to level $2$, irrespectively of the reservoir's state,
 is given at the rescaled time $t= \tp\eps$ by
\begin{equation} \label{eq-transition_proba}
  \transproba (t) = \tr \Big( \big( P_2(t) \otimes \idop \big) \U (t) \big( P_1 (0) \otimes \ketbra{\chi}{\chi} \big) \U(t)^* \Big)\;.
\end{equation}
We define the Kato unitary intertwining operator $W_K(t)$  by
\begin{equation}   \label{eq-kato_op}
\partial_t W_K(t) = K(t)W_K(t ) \; , \qquad W_K(0)=\idop\;,
\end{equation}
where
\begin{equation} \label{eq-k}
K(t)=\sum_{j=1}^2  (\partial_t P_j(t))P_j(t)=  - K^*(t)
\end{equation}
satisfies
\begin{equation}\label{pkp}
  P_j(t)K(t)P_j(t)= 0\;.
\end{equation}
The  operator $W_K(t)$ possesses the well known intertwining
property~\cite{Kato50}
\begin{equation} \label{eq-interwining_prop}
W_K(t) P_j(0)=P_j(t)W_K(t)\;.
\end{equation}
In absence of the system-reservoir coupling ($\lambda=0$) the transition probability 
(\ref{eq-transition_proba}) reduces to the adiabatic transition between the levels of a 
driven system isolated from its environment. As we shall recover along the way, the latter is known to be equal to
\begin{equation} \label{eq-transition_proba_free}
  \transprobafree (t) = \eps^2 \qtrans (t) + \Oo(\eps^3) \;,\quad
  \qtrans (t) \equiv 
  \frac{|\bra{\psi_2(0)} W_K (t)^\ast   \partial_t W_K (t) \ket{\psi_1(0)}|^2}{e_{21} (t)^2}\;.
\end{equation}  

One can motivate our choice of an instantaneous pure-dephasing model as follows. If $\HS$ and $B$ are time-independent, then
the system prepared initially in  the state $\psi_1(0)$
  remains in that state for all times.
  This mimics what happens when studying adiabatic transitions in closed systems (\ie systems uncoupled to their environment), prepared initially in an eigenstate of their Hamiltonian $\HS(0)$. If we considered a model including
  energy  exchanges such as absorption or emission processes  of a boson from the reservoir, then  transitions from one eigenstate to another induced by these processes would come into play, thus adding contributions to $\transproba(t)$ that might not vanish in the adiabatic limit and blur the adiabatic transition we are interested in.

\subsection{Reservoir autocorrelation function}

The reservoir autocorrelation function for the  zero temperature reservoir is defined by 
\begin{equation} \label{eq-gamma_def}
  \gamma(t) = 2 \bra{\chi} \E^{\I t \HR} \phi(g) \E^{-\I t \HR} \phi(g) \chi\rangle = \bra{\chi} a \big( \E^{\I t \omega} g  \big) a^\ast \big( g \big) \chi\rangle = \scalprod{\E^{\I t \omega} g}{g}\;,
\end{equation}    
where $g$ is the form factor and $\scalprod{f}{g} = \int_{\real^3} \D^3 k \,\overline{f(k)} g (k)$ stands for the scalar product in $L^2 (\real^3)$.
Assuming for concreteness a photonic dispersion relation 
$\omega(k)= | k|$, we get
\begin{equation}
	\label{2.17}
\gamma ( t)=\int_0^\infty  \D \omega\; \E^{-\I t \omega}\omega^2 \int_{S^2} \D^2 \sigma |g(\omega,\sigma)|^2,
\end{equation}    
where $g(\omega,\sigma)$ is the expression of $g(k)$ in spherical coordinates and $\D^2 \sigma$ stands for the uniform measure on the sphere $S^2$.
Hence, $\gamma(t)$ is the Fourier transform of the non-negative function
\begin{equation} \label{eq-spectral_density}
\widehat{\gamma} ( \omega)=2\pi \omega^2 \int_{S^2} \D^2 \sigma |g(\omega,\sigma)|^2 1_{\{ \omega \geq 0\}},
\end{equation}
where $1_{\{ \omega \geq 0\}}=1$ if $\omega\geq 0$ and $0$ otherwise. 
In particular, $\gamma$ is a positive definite function and thus satisfies
$\gamma ( t)=\overline{\gamma (-t)}$  and  
$|\gamma(t)|\leq \gamma(0)$.
In the physics literature $\widehat{\gamma} ( \omega)$ is also  called the power spectrum or reservoir spectral density, sometimes denoted $J(\omega)$~\cite{Weiss93}.
 If $\gamma\in L^1(\real)$, then
$\widehat{\gamma} ( \omega)= \int_\real \D t\, \E^{\I \omega t} \gamma(t)$.

 \begin{rmk} \rm
We may as well consider non relativistic massive bosons with mass $M>0$, for which $\omega(k)= |k|^2/(2M)$. Then
 \begin{eqnarray*}  
\gamma ( t) & = & \sqrt{2} M^{\frac{3}{2}} \int_0^\infty  \D \omega\; \E^{-\I \omega t }  \sqrt{\omega} \int_{S^2} \D^2 \sigma |g(\sqrt{2 M \omega},\sigma)|^2  
\\
\widehat{\gamma} ( \omega) & = & \pi  (2 M)^{\frac{3}{2}} \sqrt{\omega} \int_{S^2} \D^2 \sigma |g(\sqrt{2M \omega},\sigma)|^2 1_{\{ \omega \geq 0\}}
\end{eqnarray*}
 and the aforementioned properties of $\gamma$ still hold.

\end{rmk}

We shall make the following hypothesis, which implies in particular that  $\gamma\in L^1 (\real)$.

\begin{enumerate}[label=(A.\arabic*)]
\setcounter{enumi}{3} 

\item  \label{assum-decay_gamma} 
  $\dss \sup_{t \in \real} ( 1 + t^2)^{\frac{m+1}{2}} |\gamma(t)|  < \infty$ and 
$\dss \lim_{\omega \to 0+} \frac{\widehat{\gamma} ( \omega)}{\omega^m} \equiv \gamma_0\geq 0$ exists and is finite,  
 with $m>0$ a positive real number.
\end{enumerate}
    
These assumptions are fulfilled 
for instance for the photon dispersion relation $\omega(k)=|k|$ and for a rotation-invariant form factor $g$ of the form
\begin{equation} \label{eq-form_factor}
g(k) = g_0 |k|^{\frac{m}{2}-1} \exp \Big( - \frac{|k|}{2  \omega_D} \Big)
\end{equation}
with $m>0$, $g_0 \in \real$, and $\omega_{\rm D}>0$ a Debye cut-off frequency.
Then $\gamma(t)$ and $\widehat{\gamma} ( \omega)$ can be calculated explicitly, 
\begin{equation} \label{eq-exact_expression_gamma}
  \gamma (t) = 4\pi g_0^2 \omega_{\rm D}^{m+1} \frac{\Gamma(m+1)}{(1+\I \omega_{\rm D} t)^{m+1}} \;,\qquad
  \widehat{\gamma} ( \omega) = 8 \pi^2 g_0^2 \, \omega^m \E^{-\frac{\omega}{\omega_{\rm D}}} 1_{\{ \omega \geq 0\}}  
\end{equation}
with $\Gamma$ the Gamma function.

Let us point out that  the low-frequency behaviour  $\widehat{\gamma} ( \omega)\sim  \gamma_0\, \omega^m$ of the spectral density
determines the  time decay of the system coherences in the energy  eigenbasis
$ \{ {\psi_1}, {\psi_2}\}$ (that is, of the off-diagonal elements of the reduced density matrix
$\rho_S(t)$)  when  $\HS$ and $B$ are time-independent:  for zero temperature  reservoirs,
the decoherence is incomplete when $m>1$ (that is, the off-diagonal elements do not converge to $0$ as $t \to \infty$), whereas it is complete when $0< m \leq 1$  (the off-diagonal elements tend to zero).
The case $m>1$ is called the super-Ohmic regime, while $m=1$ and $0<m<1$ are termed the Ohmic and sub-Ohmic regimes, respectively (see e.g.~\cite{Weiss93}).

\subsection{Positive temperatures}\label{postemp1}

At positive temperatures $T=1/\beta >0$,  the reservoir equilibrium momentum distribution is given by Planck's law as  $1/(e^{\beta |k|}-1)$, where we assume that  $\omega(k)=|k|$.  Let  $\transproba(t)$ again denote the probability of transition from levels $1\rightarrow2$, where now the reservoir is initially in the temperature state. Formally, it is given by (compare with \eqref{eq-transition_proba})
\begin{equation} \label{eq-transition_proba-postemp}
	\transproba (t) = \tr \Big( \big( P_2(t) \otimes \idop \big) \U (t) \big( P_1 (0) \otimes \rho_{{\rm R},\beta} \big) \U(t)^* \Big)\;,
\end{equation}
where $\rho_{{\rm R},\beta}$ is the reservoir Gibbs density matrix. The expression \eqref{eq-transition_proba-postemp} is formal in the sense that we consider the reservoir to be infinitely extended (having continuous momentum modes), so that $\rho_{{\rm R},\beta}$ has to be interpreted as an operator in a modified Hilbert space, see Section \ref{postemp}. Another way of saying this is that in \eqref{eq-transition_proba-postemp}, we understand that the thermodynamic limit is taken, that is, we replace $\rho_{{\rm R},\beta}$ by $\rho^\Lambda_{{\rm R},\beta}$, where $\Lambda$ is a compact box in position space ${\mathbb R}^3$ (then $\rho^\Lambda_{{\rm R},\beta}$ is  a well defined operator on ${\mathcal F}_+(L^2({\mathbb R}^3))$) and we take the limit $\Lambda\nearrow{\mathbb R}^3$.

Now the Fourier transform of the reservoir autocorrelation function is
\be\label{autobeta}
\widehat\gamma^\beta(\omega)=\onehalf \widehat\gamma(|\omega|)\left(\coth(\beta | \omega |/2)+{\rm sgn}(\omega)\right)\geq 0 \;,
\ee
where $\widehat\gamma(\omega)$ is the spectral density \eqref{eq-spectral_density} and  ${\rm sgn}$ is the sign function
(see Section \ref{anpote} for a derivation of \eqref{autobeta}). As above (see \eqref{2.17}, \eqref{eq-spectral_density}), we set 
\begin{equation}
\gamma^\beta(t)=\frac{1}{2\pi}\int_{\mathbb R}\D \omega\,  e^{-i\omega t}\, \widehat\gamma^\beta(\omega)\;.
\end{equation}
We show in Section \ref{anpote} that,  in the positive temperature case, condition \ref{assum-decay_gamma} with $\gamma^\beta$, $\widehat\gamma^\beta$ in place of $\gamma$, $\widehat\gamma$, is satisfied for
  \begin{equation*} 
	\label{m19.pos}
	g(k) = g_0 |k|^{\frac{\mu}{2}-1} \exp \Big( - \frac{|k|}{2  \omega_D} \Big) \qquad \mbox{with $\mu > m+1 >1$}\;.
  \end{equation*}
Physically, the form factors we can deal with at positive temperature correspond to the super-Ohmic regime, {i.e.} $\mu >1$.

\subsection{ Main result}

Here is our main result valid for both the zero and positive temperature reservoirs. 

\begin{thm}\label{thm-1} Suppose the reservoir is initially in equilibrium at zero temperature or at temperature $1/\beta>0$. Assume that conditions
  \ref{assum-gap}-\ref{assum-decay_gamma} are satisfied for 
$\gamma$ and $\widehat\gamma$ in the former case or for $\gamma^\beta$ and $\widehat\gamma^\beta$ in the latter case and  let $m_1\equiv\min\{m,1\}>0$ and  $\alpha_0= (2m-m_1+2)^{-1}$.Then
\begin{itemize}
\item[(i)] At positive temperature $1/\beta>0$, we have 
 \begin{eqnarray} \label{eq-main_result}
    \nonumber
    \transproba (t) & = & \transprobafree (t) + \frac{\lambda^2}{2\eps} \int_0^t \D s \,\transprobafree ( s) b_{12}^2(s) \widehat{\gamma}^\beta \big( e_{12} (s) \big) 
  \\  \nonumber
  & &  +  \Oo(\eps^3) + \Oo \big(  \lambda \eps^{(3+m_1)/2} {  | \ln \eps |^{\onehalf \delta_{m,1}} } \big)
  + \Oo( \lambda^2 \eps^{1+m \alpha_0} )
  {  + \Oo \big(  \lambda^3 \eps^{(1+m_1)/2} | \ln \eps |^{\onehalf \delta_{m,1}}  \big) }
  \\
  & & 
  + \Oo( \lambda^4 \eps^{m_1} {  | \ln \eps |^{\delta_{m,1}}} ) + {  \Oo( \lambda^5 \eps^{\onehalf m_1}  | \ln \eps |^{\onehalf \delta_{m,1}} ) }+ \Oo(\lambda^6 ) ,
  \end{eqnarray}
  with $\transprobafree (t)$ the transition probability \eqref{eq-transition_proba_free} in the absence of coupling to the reservoir. (Here, $\delta_{a,b}$ is the Kronecker delta.)
 \item[(ii)] At zero temperature, the same expression \eqref{eq-main_result} holds with $\widehat\gamma$ in place of $\widehat\gamma^\beta$. 
 \item[(iii)]  In both cases, $\transprobafree (t)$ can be replaced by $\eps^2 \qtrans (t)$ in \eqref{eq-main_result}, see \eqref{eq-transition_proba_free}.
 \end{itemize}
\end{thm}

\begin{rmk} \rm \label{rmk-2}
  The theorem shows the following.
  \begin{itemize}

  \item[(i)] The second term in the \RHS of (\ref{eq-main_result}) describes, to leading order in $(\eps, \lambda)$, the modification of the transition
    probability due to the coupling with the reservoir with respect to the case without coupling. This term
    is always nonnegative, as $\widehat\gamma\geq 0$ and $\widehat\gamma^\beta\geq 0$.
     At zero temperature, it vanishes
    for transitions from the ground to the excited state, since $\widehat{\gamma}(e_{12})=0$ when $e_{12}<0$,
    see (\ref{eq-spectral_density}). 
    By contrast,  at positive temperature, we have  $\widehat{\gamma}^\beta \big( e_{12}  \big)>0$ even if $e_{12}  <0$, (assuming $|e_{12}|([0,1]) \subset \supp \hat{\gamma}$ ), see \eqref{autobeta}, so
    that the coupling with the reservoir always enhances the transition probability, be it from ground to excited state or vice versa. The
    asymmetry in the transitions to the upper and to the lower energy levels decreases with the temperature $1/\beta$, since we have
$$
\widehat\gamma^\beta(-\omega) =e^{-\beta  \omega }\, \widehat\gamma^\beta(\omega) \ \ \ \mbox{for } \omega > 0
$$
and so for large $\beta$, $\widehat\gamma^\beta(-\omega)$ decreases exponentially quickly in $\beta$.
  
\item[(ii)] To insure that the error terms be much smaller than both the first and second terms in the \RHS of (\ref{eq-main_result}), 
the coupling constant and adiabatic parameter must satisfy $\eps \ll \lambda \ll \eps^{1/3}$ when $ m > 1$ and
$ \eps^{(1+m)/2} |\ln(\eps)|^{\delta_{1,m}/2} \ll \lambda \ll \eps^{(3-m)/6}  |\ln \eps |^{-\delta_{1,m}/6}$ when $0 < m \leq 1$.
  One can further distinguish the following regimes:

\begin{enumerate}
  
\item
 If $\lambda$ scales like $\sqrt{\eps}$, the 
 transition probability 
 is larger than  its value  $\transprobafree (t)$  in absence of coupling to the reservoir   by an amount of the same order, $\eps^2$, with overall error term $o(\eps^2)$, save in the zero temperature case for the transition to the excited state.

\item   By contrast, when $\lambda \ll \sqrt{\eps}$, the system-reservoir interactions have a negligible effect on the
  transition probabilities:  $ \transproba(t) = \transprobafree(t)+o(\eps^2)$.
  
\item   For stronger coupling constants $\lambda$ such that
$\sqrt{\eps} \ll \lambda \ll \eps^{ \max{\{1/4,(1-m_1)/2\} }}$,
  the transition probability is asymptotically larger than in the absence of reservoir:
  \begin{equation} \label{eq-main_result-without_free_proba_transit}
    \transproba(t) =  \frac{\lambda^2}{2\eps} \int_0^t \D s \,\transprobafree ( s) b_{12}^2(s) \widehat{\gamma} ^\beta\big( e_{12} (s) \big) + o(\lambda^2 \eps)\;,
  \end{equation}
  save in the zero temperature case for the transition to the upper level.
  This means that the reservoir  significantly
  helps  the system to tunnel.

\end{enumerate}

\item[(iii)]    
  If  $H_S(\tau)$ is constant in a neighborhood of $t$ (but not on the whole time interval $[0,t]$), then
   $\qtrans(t)=0$ and the transition probability is given by (\ref{eq-main_result-without_free_proba_transit})
  in the wider range of coupling constants
  $ \eps^{(1+m_1)/2}  |\ln \eps |^{\delta_{m,1}/2}  \ll \lambda \ll \eps^{\max\{ 1/4, (1-m_1)/2 \}}$. Note that the integral in the \RHS of  (\ref{eq-main_result-without_free_proba_transit}) is nonzero for transitions to the lower energy state, as  $\qtrans (s)>0$
  on $[0,t]$ except for times $s$ close to $0$ and $t$.
  \end{itemize}
\end{rmk}

\begin{rmk} \rm
  The second term in (\ref{eq-main_result}) -- describing the effect of the reservoir
  on the transition probability,
  depends linearly on the adiabatic parameter $\eps$
  and quadratically on the coupling constant $\lambda$.
  A similar linear dependence on $\eps$ of adiabatic transition probabilities
  in open quantum system dynamics governed by so-called time-dependent dephasing Lindbladians  has been found in \cite{Avron10}, see also~\cite{Fraas16}. 
  Such Lindbladians share with our model the property that they instantaneously  generate pure dephasing dynamics  with no
  energy exchange. For static Hamiltonians,  they describe the evolution of the system under the Born-Markov and rotating wave approximations
  (van Hove weak coupling limit). 
  These approximations are not obvious to justify from a microscopic approach even for time-independent open systems, see {e.g.} \cite{Da, Maop}, let alone when the system Hamiltonian and the coupling depend on time.
  
  As pointed out in~\cite{Avron10,Fraas16}, the dephasing Lindbladians should be considered as phenomenological models. Although the same dependance on $\eps$ and $\lambda^2$
 (the latter corresponding to the amplitude of the dephasing dissipator of the Lindbladian) is found for both  our microscopic and the Lindbladian models,
  we stress that the Lindbladian approach does not feature any asymmetry in the transition probabilities to the upper and lower energy levels.

The papers \cite{Avron10} and \cite{Fraas16} actually consider as the system generator the emblematic Landau-Zener Hamiltonian 
$$
H_{\rm LZ}(t) = \onehalf \left( \begin{array}{cc}  t & \Delta \\ \Delta & - t \end{array} \right)\;,
$$
which gives rise, in a scattering regime, to the exact Landau-Zener formula.
This formula tells us that the transition probability is exponentially small, $\transprobafree = \E^{- \pi \Delta^2 / (2 \eps)}$, see \cite{Landau, Zener} and \cite{Joye} for generalisations. When dephasing is
included within the Lindbladian approach, the scattering limit of the transition probability is shown in \cite{Fraas16} to be
given by $\E^{- \pi \Delta^2 / (2 \eps)}$ plus an explicit term of order $\gamma_{\rm deph} \eps$, up to error terms $\Oo(\gamma_{\rm deph} \eps^2)$,
where $\gamma_{\rm deph}$ is the dephasing rate (amplitude of the dephasing dissipator).
Hence, unless the dephasing  rate is exponentially small, the Landau-Zener term is buried in the error terms.
In our approach, we consider general two-level Hamiltonians $\HS(t)$ and finite rescaled time intervals over which the explicit leading order of the transition probability $\transprobafree(t)$ is  of order $\eps^2$.
This enables us compare this contribution with that induced by the coupling to the reservoir to the full  
probability   $\transproba(t)$ in (\ref{eq-main_result}).

\end{rmk}
\begin{rmk}
\rm {\it Prima facie} the proofs of the results for zero and positive temperatures might be expected to look quite different. However, by using the so-called Gelfand-Naimark-Segal (GNS) representation of the reservoir equilibrium state at positive temperature, its density matrix is simply a rank-one projection on a vacuum state, but in a different Fock space. We explain this in Section \ref{postemp} and we show that, once the replacement of the Hilbert space is made, the proof of Theorem \ref{thm-1} for zero temperature is straightforwardly altered to accommodate for positive temperatures. 
\end{rmk}


\subsection{Organization of the paper}

The remaining part of the paper is devoted to the proof of Theorems \ref{thm-1}  and we start with the zero temperature case. In Section \ref{sec-exact_calculations} we introduce the adiabatic evolution and the corresponding wave operator and we give a Dyson series expansion of the latter. The first term in this series produces the main term in the expression for the transition probability (see \eqref{lead}). We analyze its adiabatic and weak coupling limit  in Section \ref{sec-contribution_first_term}, where the main result is Proposition \ref{prop-6}. In Section \ref{sec-higher_order_terms} we control the remaining terms in the Dyson series. The main result is Proposition \ref{prop_contibution_higher_order_terms}. We explain in Section \ref{postemp} the positive temperature formalism and the necessary changes  in the previous arguments.

\bigskip


\vspace{.3cm}
\noindent
{\bf\large  Acknowledgments} \vspace{.2cm}

This work is partially supported by the CNRS program PICS (DEASQO). We acknowledge the warm hospitality and support of the Centre de Recherches Math\'ematiques in Montr\'eal during some  stage of this work: A.J. and M.M. were supported by the
Simons Foundation and the Centre de Recherches Math\'ematiques, through the Simons-CRM scholar-in-residence program. 
A.J. acknowledges support from ANR grant NONSTOPS (ANR-17-CE40-0006-01), M.M. acknowledges support from NSERC under the Discovery Grant Programme and
D.S. acknowledges support from the Vicerrector\'{\i}a de Investigaci\'on y Desarollo de  la Universidad de Concepci\'on, proyecto 
VRID 218.013.045-1.OIN, and the Fondecyt project N$^0$ 1190134.

\section{Exact calculations and adiabatic Dyson expansion} \label{sec-exact_calculations}

\subsection{Expansion of the wave operator}

Let us consider the adiabatic evolution operator $\UA (t)$ solution of 
\begin{equation} \label{eq-Shrodinger_eq_U_ad}
  \I \eps \partial_t \UA (t) = \big( H(t) + \I \eps K(t) \otimes 1 \big) \UA (t) \; , \qquad \UA (0 ) = \idop\;, \qquad  t \in (0,1)\; ,
\end{equation}  
where $K(t)$ is given by (\ref{eq-k}) and $H(t)= \HS (t) \otimes \idop + \Hint (t)  + \idop \otimes \HR$ is the total Hamiltonian of the system and reservoir. Since $\I \eps K(t)$ is self-adjoint, the operators $\UA (t)$ are unitary.
For later purposes, we define the unitary dynamical phase operator $\UPsi(t)$ given by 
\begin{equation} \label{eq-definition_dynamical_phase_op} 
  \UA (t) = (\WK (t) \otimes \idop) \ \UPsi (t)\;.
\end{equation}
 The justification of this designation will be provided by Lemma~\ref{lem-1}, which shows that $\UPsi(t)$ can be computed exactly, commutes with $P_j(0)\otimes \idop$ and contains fast-oscillating terms as $\eps\to 0$. 
  
Before showing that, let us introduce the unitary  wave operator 
\begin{equation} \label{eq-wop}
\Waveop (t)=\UA^*(t) \U (t)\; , \quad  t \in [0,1]\;.
\end{equation}
In view of (\ref{eq-Schrodinger_Ueps}), (\ref{eq-Shrodinger_eq_U_ad}), and $K (t)^\ast = - K(t)$, this operator satisfies
$ \partial_t \Waveop (t)=-\UA^*(t) (K(t)\otimes \idop ) \UA (t) \Waveop (t)$ and $\Waveop (0)= \idop$ or, equivalently,
\begin{equation} \label{eq-edomega}
  \partial_t \Waveop (t)=- \UPsi (t)^\ast \widetilde{K} (t) \UPsi (t) \Waveop (t)
  \; ,  \qquad \Waveop (0)=\idop\;,
\end{equation}
where $\widetilde{K} (t) =  \WK^*(t) K(t) \WK(t)\otimes \idop$ is independent of $\eps$ and acts trivially on the reservoir.
Note that the dependence of $\Waveop (t)$ on $\eps$ comes from the fast-oscillating factors in the dynamical phase operator only.

Upon substituting the \RHS of (\ref{eq-edomega}) into $\Waveop (t) = \idop + \int_0^t \D s \, \partial_s \Waveop (s)$ and iterating, one obtains
the norm-convergent Dyson expansion
\begin{align}\label{eq-dyson}
\Waveop (t)&= \sum_{k\geq 0} \Waveop^{(k)}(t) \\ \nonumber 
&=\sum_{k\geq 0} (-1)^k \! \int \!\! \ldots \!\! \int_{0 \leq s_k \leq \cdots \leq s_1 \leq t} \!\!\!\!\!\! \D s_1  \cdots \D s_k \,
 \UPsi^\ast (s_1) \widetilde{K} (s_1) \UPsi (s_1) \cdots \UPsi^\ast (s_k) \widetilde{K} (s_k) \UPsi (s_k) 
\end{align}
such that for $k\geq 1$
\begin{equation} \label{eq-oj}
\Waveop^{(k)} (t)=- \int_0^t \D s\, \UPsi^\ast (s) \widetilde{K}  (s)  \UPsi (s) \Waveop^{(k-1)} (s)\; , \quad \Waveop^{(0)} (t)=\idop\; .
\end{equation}
We may now rewrite the transition probability (\ref{eq-transition_proba}) in terms of $\Waveop (t)$ by proceeding as follows
\begin{eqnarray} \label{eq-transition_proba_in_terms_waveop}
\nonumber  
  \transproba (t)
  & = & \tr \Big[ P_2 (t) \UA (t) \Waveop (t) ( P_1(0) \otimes \ketbra{\chi}{\chi} ) \Waveop (t)^\ast \UA (t)^\ast \Big]
  \\ \nonumber
  & = & \tr \Big[ \WK (t)^\ast P_2 (t) \WK (t)  \UPsi (t) \Waveop (t) ( P_1(0) \otimes \ketbra{\chi}{\chi} ) \Waveop (t)^\ast \UPsi (t)^\ast \Big]
  \\
  & = & \big\| P_2 (0)  \Waveop (t)  \psi_1 (0) \otimes \chi \big\|^2\;,
\end{eqnarray}
where the intertwining property (\ref{eq-interwining_prop}) and the commutation of the dynamical phase operator with the projectors $P_j(0)$ have been used to get the last expression.

Introduce now for $i,j\in\{1,2\}$
\begin{equation} \label{eq-def_tilde_K}
\widetilde K_{ij}(t) \equiv P_i(0)\widetilde K(t) P_j(0)= W_K^*(t)P_i(t)K(t)P_j(t)W_K(t)
\end{equation}
and observe that $\widetilde K_{jj}(t)= 0$ (due to (\ref{pkp})). Using once again the commutation of the dynamical phase operator
  with the projectors $P_j(0)$, this implies, (dropping  $\otimes \idop$ from the notation), 
\be\label{kofdiag}
\Waveop^{(k)} (t) P_1(0) =\left\{\begin{matrix} P_2(0) \Waveop^{(k)} (t) P_1(0) & \mbox{for}\ \ k \ \ \mbox{odd,}\cr
P_1(0) \Waveop^{(k)} (t) P_1(0) & \mbox{for}\ \ k \ \ \mbox{even.}\end{matrix}\right.
\ee
Hence 
the substitution of the series (\ref{eq-dyson}) into (\ref{eq-transition_proba_in_terms_waveop}) yields
\begin{align} \label{eq-trans_proba_infinite_sum}
\transproba (t)&=\left\| P_2(0)  \sum_{k \ \mbox{\tiny odd}}\omeg{k} (t) \right\|^2 =
\sum_{ {j \ \mbox{\tiny odd} }\atop{k \ \mbox{\tiny odd} } } \langle \omeg{j} (t) | \omeg{k} (t) \rangle
\end{align}
with 
\begin{equation} \label{eq-def_omega_k}
   \omeg{k} (t) \equiv \Waveop^{(k)} (t) \, {\psi_1(0)} \otimes \chi  \;.
\end{equation}
Thus
\begin{align}\label{lead}
&\Big|\transproba (t)-\| \omeg{1} (t)\|^2 \Big| \leq 
2 \| \omeg{1} (t)\|  \sum_{k \geq 3 , \ k \ \mbox{\tiny odd}}\| \omeg{k} (t)\| +
\bigg( \sum_{k \geq 3,\ k \ \mbox{\tiny odd}}\| \omeg{k} (t) \| \bigg)^2. 
\end{align}
Therefore, if one is able to show that
\be \label{eq-sum_higher_term_Dyson}
 \sum_{k \geq 3, \ k \ \mbox{\tiny odd}}\| \omeg{k} (t) \|\;\; \ll \;\; \| \omeg{1} (t) \| \; ,
\ee
it will follow that $\transproba (t) = \|  \omeg{1}(t) \|^2+o(\| \omeg{1}(t) \|^2)$.
This is what we set out to prove (see Sec.~\ref{sec-higher_order_terms} below).

\subsection{Exact expression of the dynamical phase operator}

For our free boson reservoir model, an exact expression of $\UPsi (t)$
can be obtained in terms of the Weyl operators.
Recall that the latters are unitary operators on $\mathcal{F}_{+}(L^2(\real^3))$, defined by
 $W(F)=\E^{\I \phi(F)}$ for any $F \in L^2 (\real^3)$ (see~\cite{Bratteli} for more details).

\begin{lem} \label{lem-1}
  Let $\varphi_{j} (t) = {\eps}^{-1}  \int_0^t \D s \, e_j (s)$, $j=1,2$ denote the dynamical phases of the system Hamiltonian $\HS$.
  Then, the dynamical phase operator defined by (\ref{eq-definition_dynamical_phase_op}) is given by 
\begin{equation} \label{eq-Psi}
  \UPsi (t) = \sum_{j=1}^2 \E^{-\I \varphi_{j}(t) }  P_j(0) \otimes \E^{-\frac{\I t }{\eps} \HR} \X{j} (t)\;,
\end{equation}
where the  unitary operators $\X{j} (t)$ on $\mathcal{F}_{+}(L^2(\real^3))$ are given by
\begin{equation} \label{eq-Xj}
  \X{j} (t) = \E^{\I\, \zeta_{j} (t)}   W(F_{j} (t))\; , \qquad j=1,2
\end{equation}
with
\begin{eqnarray} 	\label{eq-zetaj_Fj}
\nonumber F_j(t) & = & \int_0^t  \D s \, f_{j}(s) \\ 
\zeta_{j} (t) & = & \onehalf \ \im \int_0^t \D s \, \scalprod{F_j(s)}{f_j(s)} = \onehalf \ \im  \int_0^t \D s \int_0^s \D \tau \, \scalprod{f_{j} (\tau)}{f_{j} (s)} 
\end{eqnarray}
and
\begin{equation} \label{eq-fj}
f_j(t ) = -\frac{\lambda}{\eps} \,b_j(t)\E^{\I\omega(\cdot ) t/\eps} g(\cdot )  \in  L^2({\mathbb R}^3).
\end{equation}
\end{lem}

To simplify notation we do not indicate the dependence of $\varphi_{j}(t)$, $\zeta_j (t)$, $F_{j} (t)$, $f_j(t)$,
and $\X{j}(t)$ on $\eps$ and $\lambda$.

\begin{rmk} \rm
The operator $\UPsi (t)$ is diagonal in the eigenbasis of $\HS(0)$ and in absence of coupling to the reservoir 
$P_j(0) \Psi_{0,\eps}(t)=\E^{-\frac{\I}{\eps} \int_0^t \D s \, e_j (s)} P_j(0)$ coincides with the dynamical phase, see~\cite{Kato50}.
For $\lambda>0$, $P_j(0) \UPsi (t)$ has a
non-trivial action on the reservoir degrees of freedom and contains other fast-oscillating factors 
depending on the interaction and reservoir Hamiltonians $\Hint (t)$ and  $\HR$.
\end{rmk}

\vspace{2mm}

\Proof Plugging  (\ref{eq-definition_dynamical_phase_op}) into (\ref{eq-Shrodinger_eq_U_ad}) and taking advantage of (\ref{eq-kato_op}),  (\ref{eq-interwining_prop}) and (\ref{H_S}), one finds that
$\UPsi (t)$ satisfies 
\begin{eqnarray} \label{eq-Schrodinger_eq_for_Psi}
  \nonumber
  \I \eps \partial_t \UPsi (t) & = & \WK (t)^\ast H(t)\WK(t)  \,\UPsi (t)\\
  & = & \sum_j  P_j (0) \otimes \Big( e_j(t ) \idop + \lambda b_j(t) \phi (g) +  \HR \Big) \UPsi (t)
  \;,\quad \UPsi (0)= \idop\;.
\end{eqnarray}
The solution of this equation is given by (\ref{eq-Psi}) with $\X{j} (t)$ the unitary operators   on $\mathcal{F}_{+}(L^2(\real^3))$
given by
\begin{equation}\label{m17}
  \I \eps \partial_t \X{j} (t) = \lambda b_j(t) \E^{\frac{\I t }{\eps} \HR} \phi (g) \E^{-\frac{\I t }{\eps} \HR} \X{j} (t)\; , \qquad \X{j} ( 0) = \idop\; .
\end{equation}  
The expression of $\X{j} (t)$ in terms of the Weyl operators in the lemma is obtained
from this equation, using also $\E^{\frac{\I t }{\eps} \HR} \phi (g) \E^{-\frac{\I t }{\eps} \HR}= \phi ( \E^{\I\omega t/\eps} g  )$
and the commutation relation~\cite{Bratteli} 
\begin{equation} \label{eq-Wphi}
[ W( F)\, , \, \phi( G) ] = -\im  \{ \scalprod{F}{G}) \} W( F)\;,\qquad \forall\;\;F, G \in L^2 (\real^3)\;,
\end{equation}
which yields
\begin{eqnarray} \label{eq-derweyl}
  \partial_t W( F(t)) \nonumber
  & = & \I \int_0^1 \D u \,\E^{\I u \phi (F(t))} \phi ( \partial_t F (t)) \E^{\I (1-u) \phi (F(t))} \\
  & = & \Big( \I \phi(\partial_t F )-\frac{\I}{2}  \im  \scalprod{F(t)}{\partial_t F(t)}  \Big) W(F(t))\;.
\end{eqnarray}
\hfill{\finpro}

\subsection{Iterative formula for the vectors $\omeg{k}$}
%
Using  the property of the Weyl operators
\begin{equation} \label{eq-product_Weyl_op}
  W( f ) W( g ) = \E^{-\frac{\I}{2} \im \scalprod{f }{g}} W ( f + g) \; , \qquad f, g \in L^2 (\real^3)\;,
\end{equation}
one infers  from (\ref{eq-oj}), (\ref{eq-def_tilde_K}), (\ref{kofdiag}),  (\ref{eq-def_omega_k}),  and Lemma~\ref{lem-1} that
\begin{equation} \label{eq-iterative_formula_omega_k_k-1}
  {\omeg{k}(t)} =
  \begin{cases} \displaystyle
    - \int_0^t \D s\, \E^{-\I ( \dynphase{1}{2} (s) - \zeta_{12} (s) + \frac{1}{2} \im \scalprod{F_1(s)}{F_2(s)} )} \Kt{2}{1} ( s) \otimes W ( F_{12}(s) )  {\omeg{k-1}(s)} & \text{ if $k$ is odd}
 \\[3mm]   \displaystyle
   - \int_0^t \D s \, \E^{-\I ( \dynphase{2}{1} (s ) - \zeta_{21} (s)  + \frac{1}{2} \im \scalprod{F_2(s)}{F_1(s)} )} \Kt{1}{2} ( s) \otimes W ( F_{21}(s) )  {\omeg{k-1}(s)} & \text{ if $k$ is even,}  
  \end{cases}
\end{equation}
where we have introduced the dynamical Bohr frequencies 
\begin{equation} \label{eq-dyn_phase}
  \dynphase{i}{j} ( t) \equiv  \varphi_{i}(t) - \varphi_{j} (t) = \frac{1}{\eps} \int_0^t \D u \, \big( e_i (u) - e_j (u) \big)\;, \qquad i \not= j \in \{ 1,2\} 
\end{equation}  
and  we have set similarly $\zeta_{ij}(t) \equiv \zeta_i(t)-\zeta_j(t)$ and $ F_{ij} (t) \equiv F_{i} (t) - F_{j}(t)$.

An iteration formula for ${\omeg{k}(t)}$ for  odd $k$'s, $k \geq 3$, is obtained by
plugging the second equation into the first one in (\ref{eq-iterative_formula_omega_k_k-1}). Using (\ref{eq-product_Weyl_op}) again, this yields
  ($k \geq 3$, $k$ odd)
\begin{equation} \label{eq-iterative_formula_omega_k}
 {\omeg{k}(t)} = \int_0^t \D s \int_0^s \D \tau \, \E^{-\I (\dynphase{1}{2} - \zeta_{12}+ \theta_{12}^{-} ) (s,\tau)} \Kt{2}{1} (s) \Kt{1}{2} (\tau) \otimes W ( F_{12}(s,\tau) ) {\omeg{k-2}(\tau)}  \;,   
\end{equation}
where
\begin{equation} \label{eq-theta12}
  \theta_{12}^\pm (s,\tau) \equiv 
  \onehalf \im \big\{ \scalprod{F_1(s)}{F_2(s)} -  \scalprod{F_1(\tau)}{F_2(\tau)}  \pm \scalprod{F_{12}(s)}{F_{12}(\tau)} \big\}
   = - \theta_{12}^\pm (\tau,s) = - \theta_{21}^\mp (s,\tau) 
\end{equation}
and we have introduced the notation 
\begin{equation} \label{eq-def_h(s,tau)}
 h_{ij} (s,\tau) \equiv h_{ij} (s)-h_{ij} (\tau) \quad \text{ for }\quad h= \varphi, \zeta, F , \text{ etc.}
\end{equation}
%

\section{Contribution of the first term in the Dyson expansion} \label{sec-contribution_first_term}

\subsection{Integration by parts}

We now use the identity
\begin{equation} \label{eq-expectation_Weyl}
  \bra{\chi} W ( f) \chi\rangle = \E^{-\| f\|^2/4} \; , \qquad f \in L^2 (\real^3)
\end{equation}
to obtain an exact expression of the main term  $ \| \omeg{1} (t) \|^2$ in (\ref{lead}).
The latter is given by (\ref{eq-iterative_formula_omega_k_k-1}) with ${\omeg{0} (t)} = {\psi_1(0)} \otimes \chi$.
Taking advantage of the antisymmetry of $\varphi_{12}$, $\zeta_{12}$, and $\theta_{12}^{\pm}$ under the exchange of
$s$ and $\tau$, a simple calculation yields
\begin{equation} \label{eq-exact_formula_norm_omega_1}
  \| \omeg{1} (t) \|^2 = 2 \re \left\{
  \int_0^t \D s \int_0^s \D \tau\,\E^{- \I \dynphase{1}{2} (s,\tau)+\I \zeta_{12} (s,\tau)-\eta_{12}(s,\tau)}
    e_{21}(\tau)^2 \qtrans (s,\tau) \right\} 
\end{equation}  
 with  $\dynphase{1}{2} (s,\tau)$, $\zeta_{12} (s,\tau)$ defined in  (\ref{eq-dyn_phase}), (\ref{eq-def_h(s,tau)}) and
  \begin{eqnarray} \label{eq-def_qtrans_eta}
    \nonumber
    \qtrans (s,\tau) & = & \frac{\bra{\psi_1 (0)} \Kt{2}{1} ( \tau)^\ast \Kt{2}{1} (s) {\psi_1 (0)}{\rangle}}{e_{21} (\tau)^2}\\
    \eta_{12}(s,\tau) & = &  \frac{1}{4} \| F_{12} (s,\tau) \|^2 + \I \theta_{12}^+ (s,\tau) \;.
  \end{eqnarray}
  Observe that by the gap and smoothness assumptions~\ref{assum-gap} and~\ref{assum-smoothness},
  $\qtrans (s,\tau)$ and its first three derivatives are bounded uniformly by
\begin{equation} \label{eq-uniform_bound_q}
  q_{\infty}^{(n)} \equiv \sup_{ 0\leq \tau \leq s \leq 1} | \partial^n_\tau \qtrans (s , \tau) |\leq c_n 
   \max_{ \nu=0,\ldots, n+1}\sup_{ 0\leq \tau\leq 1}\| \partial_\tau^\nu P_1 (\tau)\|, \ \ 0\leq n\leq 3,
\ee
where the positive constant $c_n$ depends on $\max_{ \nu=0,\ldots n}\sup_{0\leq \tau\leq 1}|\partial_\tau^\nu e_{21}|$ {and $\delta$. In what follows, we write $q_{\infty}\equiv q_{\infty}^{(0)}$}.

  Our main tool to estimate the \RHS of (\ref{eq-exact_formula_norm_omega_1}) is the following 
integration by part formula.

\begin{prop} \label{lem-2}
  Under assumptions~\ref{assum-gap}-\ref{assum-constant_eigenvectors}, one has
  \begin{eqnarray}  \label{eq-IPP_formula_first_term}
    \| \omeg{1} (t) \|^2   \nonumber
    & = &
    \eps^2\qtrans (t,t) - 2 \eps^2 \re \int_0^t \D s \int_0^s \D \tau\, \E^{- \I  \dynphase{1}{2} (s,\tau)} 
    \\
    & & \times \partial_\tau \bigg(  \frac{1}{e_{21}(\tau)} \partial_\tau
      \Big( \E^{( \I  \zeta_{12} - \eta_{12}) (s, \tau)} e_{21}(\tau) \qtrans (s,\tau) \Big) \bigg)  \;.
 \end{eqnarray}
\end{prop}


\vspace{2mm}

\Proof
The statement follows by  integrating by parts twice the $\tau$-integral in (\ref{eq-exact_formula_norm_omega_1}),
using $e_{21} (\tau) \E^{-\I \dynphase{1}{2}(s,\tau)}= \I \eps  \partial_\tau ( \E^{-\I \dynphase{1}{2} (s,\tau)})$. Noting that
\begin{enumerate}[label=(\roman*)]
\item \label{item-1_prof_lemma2}
{ $ \dss \lim_{t \to 0+} \partial^n_t K (t) =  \lim_{t \to 0+} \partial^n_t \widetilde K_{ij}(t) =   0$ for any $n=0, \ldots,3$ and $i,j=1,2$,}
  by  assumption~\ref{assum-constant_eigenvectors}, (\ref{eq-kato_op}) and (\ref{eq-k}), and (\ref{eq-def_tilde_K});
\item  \label{item-2_prof_lemma2} $\dynphase{1}{2} (s,s)= \zeta_{12}( s,s)= \eta_{12}(s,s) = 0$,
\end{enumerate}%
we get $(\partial^n_\tau \qtrans ) (s,0 )=0$ for $n\in\{0,\dots,3\}$ and 
the boundary term in the first integration by parts is equal to $\I \eps e_{21}(s) \qtrans (s,s) \in \I \real$ which disappears after taking the real part.
We arrive at 
\be
 \| \omeg{1} (t) \|^2 = - 2 \eps \re \left\{ \I 
  \int_0^t \D s \int_0^s \D \tau\,\E^{- \I \dynphase{1}{2} (s,\tau)}\partial_\tau \left(\E^{\I \zeta_{12} (s,\tau)-\eta_{12}(s,\tau)}
    e_{21}(\tau) \qtrans (s,\tau)\right) \right\} \;.
\ee
The contribution of the boundary term in the second integration by parts is estimated as follows:
\begin{align*}
& 2 \eps^2\  \re \bigg\{ \int_0^t \D s\,\bigg[\frac{ \E^{-\I \dynphase{1}{2} (s,\tau)} }{e_{21}(\tau )}
\partial_\tau \big( \E^{(\I  \zeta_{12} - \eta_{12})(s, \tau)} e_{21}(\tau) \qtrans (s,\tau) \big) \bigg]_{\tau=0}^{\tau=s} \bigg\}
\\
& = 2 \eps^2 \int_0^t \D s\, \bigg( - \re \{ \partial_\tau \eta_{12} \} (s,s) \qtrans (s,s)
  + \frac{\partial_\tau e_{21} (s)}{e_{21}(s)}  \qtrans (s,s)  + \re \{\partial_\tau \qtrans \}  (s, s) \big\} \bigg)
\\
& = 2 \eps^2 \int_0^t \D s\, \frac{d}{ds} \Big( \frac{\qtrans (s,s)}{2} \Big)   = \eps^2 \qtrans (t,t) \;.
\end{align*}
In the third equality, we have used that $\re\{ \partial_\tau \eta_{12} \} (s,s) =0$, which 
follows from differentiating
\begin{equation}
  \| F_{12} (s,\tau) \|^2 = \left\| \int^s_\tau \D u\, f_{12}(u) \right\|^2
  = \frac{\lambda^2}{\eps^2} \int_\tau^s \D u \int_\tau^s \D v \,b_{12}(u) b_{12} (v) \gamma \Big( \frac{u-v}{\eps} \Big)\;,
\end{equation}
see~(\ref{eq-gamma_def}) and (\ref{eq-fj}). The  integral term of the second  integration by parts gives rise to the double integral in
(\ref{eq-IPP_formula_first_term}). \hfill{\finpro}  

\vspace{4mm}

In the absence of coupling with the reservoir, $\E^{(\I \zeta_{12}-\eta_{12})(s,\tau)}\equiv 1$ and another integration by parts shows that
the double integral in (\ref{eq-IPP_formula_first_term}) is  of order $\eps^3$ or smaller.
When $\lambda >0$ however, the  integral term after such a third integration by parts
is not small, due to the presence of the third derivatives of $\zeta_{12}$ and $\eta_{12}$ that make
 factors $1/\eps$ appear upon differentiating $f_j(t)$ in (\ref{eq-fj}),
It is then necessary to analyze more carefully the different contributions coming from the
first and second derivatives of the $\eps$-dependent exponential $\E^{(\I \zeta_{12}-\eta_{12})(s,\tau)}$. This will be done in Subsection~\ref{sec-behavior_first_term_Dyson}. We will show that the second derivative $\partial_\tau^2 \eta_{12}(s,\tau)$ yields a contribution  
$\Oo(\lambda^2 \eps)$ 
in (\ref{eq-IPP_formula_first_term}), while the other derivatives yield much smaller contributions in the limit
$\eps \ll 1$, $\lambda \ll \eps^\alpha$ with  $\alpha \in (0, 1/2)$
some fixed exponent. 
Our analysis is based on preliminary estimations of integrals involving the derivatives of  $\eta_{12}$ and $\zeta_{12}$,
which are spelled out in the next subsection.

\subsection{Estimations on the derivatives of $\eta_{12}$  and $\zeta_{12}$} \label{sec-estim_eta_zeta}
\subsubsection{Preliminaries} \label{sec-preliminaries_estim_eta_zeta}

It is convenient to rewrite the expression on the \RHS of (\ref{eq-def_qtrans_eta}) as 
\begin{equation*} 
  \eta_{12}(s,\tau)  =  
  \frac{1}{4} \scalprod{F_{12}(s,\tau)}{F_{12}(s,\tau)} + \frac{\I}{2} \im \scalprod{F_{12}(s,\tau)}{F_1(s)}
   -  \frac{\I}{2} \im \scalprod{F_{2}(s,\tau)}{F_{12}(\tau)}\;.
\end{equation*}
Let us set $\gamma_R (x)=\re \gamma(x)$ and $\gamma_I (x)= \im \gamma (x)$. By
 (\ref{eq-gamma_def}), (\ref{eq-zetaj_Fj}), and (\ref{eq-fj}), we get 
\begin{eqnarray} \label{eq-exact_expression_eta}
\nonumber  \eta_{12}(s,\tau)   & = &
   \frac{\lambda^2}{2\eps^2} \int_\tau^s \D u \bigg[ \onehalf \int_\tau^s \D v\, b_{12}(u) b_{12} (v) \gamma_R \Big( \frac{u-v}{\eps} \Big)
     + \I \int_0^s \D v \, b_{12} (u) b_1 (v) \gamma_I \Big( \frac{u-v}{\eps} \Big)
     \\
   & &   - \I \int_0^\tau \D v \, b_{2} (u) b_{12} (v) \gamma_I \Big( \frac{u-v}{\eps} \Big) \bigg]\;.
\end{eqnarray}
Similarly, 
\begin{equation} \label{eq-exact_expression_zeta}
  \zeta_{12}(s,\tau) = - \frac{\lambda^2}{2 \eps^2} \int_\tau^s \D u \int_0^u \D v \big( b_1(u) b_1(v) - b_2(u) b_2 (v) \big)
   \gamma_I  \Big( \frac{u-v}{\eps} \Big)\;.
\end{equation}

At first sight, $\eta_{12}(s,\tau)$, $\zeta_{12}(s,\tau)$, and their first derivatives with respect to $\tau$  seem to be of order $\lambda^2/\eps^{2}$,
while their second derivatives  seem to be of order $\lambda^2/\eps^{3}$.
This would imply that  the factor $\eps^2$ gained upon integrating by parts in lemma~\ref{lem-2} is lost because of the fast oscillations and damping in
the integral induced by the reservoir. Actually, this is  not true for regular enough  form factors: as we shall prove below,
$\partial_\tau \eta_{12}$ and $\partial_\tau \zeta_{12}$ turn out to be,  after suitable integrations over $\tau$, of order $\lambda^2 \eps^{\min\{ m-1,0\}}$ and $\lambda^2/\eps$, respectively, while
$\partial_\tau^2 \eta_{12}$ and $\partial_\tau^2 \zeta_{12}$ are of order $\lambda^2/\eps$.

It is clear from the formulas above that $\eta_{12}, \zeta_{12}$ and their derivatives
depend essentially on integrals of the real and imaginary parts of the reservoir  autocorrelation function $\gamma$.  
The crucial property that we will use below, which follows from  assumption~\ref{assum-decay_gamma}, is that  $\gamma \in L^1(\real )$ and
\begin{equation}  \label{eq-vanishing_int_gamma_R}
  \int_0^\infty \gamma_R(x) \D x  =0\;.
\end{equation}  
Indeed, \ref{assum-decay_gamma}  implies that
\begin{equation*}
0=\widehat{\gamma} (0)= \int_\real \gamma_R(x) \D x + \I \int_\real \gamma_I(x) \D x = 2 \int_0^\infty \gamma_R(x) \D x,
\end{equation*}
since $\gamma_R$ and $\gamma_I$ are even and odd integrable functions, respectively.

\subsubsection{Estimations on the derivatives of $\eta_{12}$}

\begin{prop} \label{prop-1}
Suppose that $\gamma \in L^1 (\real)$ and that (\ref{eq-vanishing_int_gamma_R}) and \ref{assum-smoothness} hold true and set
\begin{equation} \label{eq-def_d(z)}
    r (z) \equiv \int_0^{z} \D y \int_y^\infty \D x \,| \gamma(x) | + \int_0^{z} \D x \, x | \gamma (x) |  \;, \ \ z\geq 0.
  \end{equation}
Let $h(s,\tau)$ be a continuous function on $[0,1]^2$, which may depend on $\eps$ and $\lambda$, bounded uniformly by an $(\eps,\lambda)$-independent constant
  $N$,   $\sup_{0 \leq \tau \leq s \leq 1} | h (s,\tau) | \leq N < \infty$. Then
 there exists a constant $c  <\infty$ independent of $\lambda$ and $\eps$ such that  for any $0 < s \leq t \leq 1$, the following bounds hold:
\begin{equation}
   \sup_{\tau, s  \in [0,t]} | \eta_{12} (s,\tau) | \leq  c \lambda^2 r \Big( \frac{t}{\eps} \Big)\;,
\end{equation}
  \begin{equation}
\int_0^t \D \tau\, | \partial_\tau \eta_{12} (s,\tau) | \leq   c \lambda^2  r \Big( \frac{t}{\eps}  \Big)
    \ , \quad
    \int_0^t \D \tau\, | \partial_\tau \eta_{12} (s,\tau) |^2 \leq   c  \frac{\lambda^4}{\eps}  r \Big( \frac{t}{\eps}  \Big)\;,
  \end{equation}
  and
  \begin{eqnarray}
  \nonumber  
  &&   \bigg| \int_0^s \D \tau \, h(s,\tau) \partial^2_\tau \eta_{12} (s,\tau)
  -\frac{\lambda^2}{2 \eps}  \int_0^{\frac{s}{\eps}} \D x \,\Big( h(s,s-\eps x)  b_{12} (s)^2 \gamma (-x)
  - \I h(s, \eps x) b_{12}(0)^2  \gamma_I (x) \Big) \bigg|
       \\
    && \hspace*{1cm}   
    \quad \leq \quad  c \lambda^2  r \Big( \frac{s}{\eps}  \Big) .
  \end{eqnarray}  
  \end{prop}

\begin{cor} \label{eq-corollary-bound_int_derivative_q}
  Suppose that assumptions~\ref{assum-smoothness} and ~\ref{assum-decay_gamma} hold and let
  us set $m_1 \equiv \min \{ m ,1\}$. Let $ 0 \leq t  \leq 1$. If $m \neq 1$, then
  \begin{equation*}
    \int_0^t \D s \int_0^t \D \tau \, | \partial_\tau \eta_{12} ( s, \tau)|  \leq  c \lambda^2 \eps^{m_1-1}
    \quad, \quad
    \int_0^t \D s \int_0^t \D \tau \, | \partial_\tau \eta_{12} ( s, \tau)|^2 \leq c \lambda^4 \eps^{ m_1 -2}
    \;,
  \end{equation*}
and if $m=1$, then
  \begin{equation*}
    \int_0^t \D s \int_0^{t'} \D \tau \, | \partial_\tau \eta_{12} ( s, \tau)|  \leq  c \lambda^2 |\ln \eps |
    \quad, \quad
    \int_0^t \D s \int_0^{t'} \D \tau \, | \partial_\tau \eta_{12} ( s, \tau)|^2 \leq c \lambda^4 \frac{| \ln \eps|}{\eps}
    \;,
  \end{equation*}

where $c$ is a constant  independent of $\lambda$, $\eps$, $t'$ and $t$.
\end{cor}  

\Proofof{Corollary~\ref{eq-corollary-bound_int_derivative_q}}
By  Assumption~\ref{assum-decay_gamma} one has $|\gamma(x) | \leq \kappa  x^{-m-1}$ for $x\geq 1$, with $\kappa$ a positive constant.
An explicit calculation then  shows that for any $0 \leq t \leq 1$,
  \begin{equation} \label{eq-bound_on_r}
    r\Big( \frac{t}{\eps} \Big) \leq r\Big( \frac{1}{\eps} \Big)
    \leq
    \begin{cases}
      r_m & \text{if $m>1$} \\
      r_1 | \ln \eps | & \text{if $m=1$}\\
      r_m \eps^{m-1}  & \text{if $0<m<1$}
    \end{cases}
  \end{equation}  
with $r_m$ a positive finite constant independent of  $\lambda$, $\eps$ and $t$.
\hfill{\finpro}

\vspace{3mm}

Before proving the Proposition, let us discuss a Corollary  which gives rise to the second term in our formula (\ref{eq-main_result}) for the transition probalility.
Consider the term  obtained by spelling out the $\tau$-derivatives in  formula (\ref{eq-IPP_formula_first_term}) and keeping only the term involving $\partial_\tau^2 \eta_{12}(s,\tau)$
\begin{equation} \label{eq-def_J_12}
J_{1 \rightarrow 2}^{(\lambda,\eps)} (t)  \equiv  
    2 \eps^2 \re  \int_0^t \D s \int_0^s \D \tau\, \E^{ (- \I \dynphase{1}{2}+\I \zeta_{12} -\eta_{12}) (s,\tau)} \partial_\tau^2 \eta_{12} (s,\tau) \qtrans (s,\tau) \;.
\end{equation}

\vspace{2mm}

\begin{cor} \label{corr-corollary-estimation_J_12}
  Suppose  that assumptions~{ \ref{assum-gap}-\ref{assum-decay_gamma}} hold. Then
  \begin{equation}
    J_{1 \rightarrow 2}^{(\lambda,\eps)} (t) =  \frac{\lambda^2 \eps}{2} \left( \int_0^t \D s\, b_{12}(s)^2 \qtrans (s,s) \widehat{\gamma} ( e_{12}(s) )
    + \Oo( \eps^{m\alpha} ) + \Oo(\eps^{1-2\alpha} ) + \Oo(\lambda^2 \eps^{-\alpha} ) \right)\;,
  \end{equation}
  where the exponent $\alpha$ can be chosen arbitrarily in $(0,1/2)$.
\end{cor}  

\Proofof{Corollary~\ref{corr-corollary-estimation_J_12}}
In view of (\ref{eq-uniform_bound_q}) and $\re \eta_{12}(s,\tau) \geq 0$, the function $h =  \E^{- \I  \dynphase{1}{2}+\I \zeta_{12} -\eta_{12}} \qtrans$ satisfies the hypotheses of 
Proposition~\ref{prop-1} with $N= q_\infty$. Thus  if $m\neq 1$,
\begin{eqnarray} \label{eq-J_12}
  \nonumber
 & &  J_{1 \rightarrow 2}^{(\lambda,\eps)} (t)
    = 
   \lambda^2 \eps \re \bigg\{
   \int_0^t \D s \bigg(  b_{12} (s)^2  \int_0^{\frac{s}{\eps}} \D x \, \E^{(- \I  \dynphase{1}{2}+\I \zeta_{12} -\eta_{12}) (s,s-\eps x)}
    \qtrans (s,s-\eps x) \gamma (-x)
   \\ 
   & & \hspace*{5mm} - \I  b_{12} (0)^2   \int_0^{\frac{s}{\eps}} \D x \, \E^{ (- \I  \dynphase{1}{2}+\I \zeta_{12} -\eta_{12}) (s, \eps x)}
    \gamma_I (x) \qtrans (s , \eps x) \bigg) \bigg\} + O ( \lambda^2 \eps^{1+m_1}) \;.
\end{eqnarray}
For $m=1$, the same result holds with an  error term of order $\lambda^2 \eps^2 | \ln \eps|$.
Now, recalling that $\qtrans (s,0) =0$ by  assumptions~\ref{assum-constant_eigenvectors}, one has
$| \qtrans (s,\eps x)| \leq q_\infty^{ (1)} \eps x$, with $q_\infty^{ (1)}$ given by (\ref{eq-uniform_bound_q}).
Hence the last integral in (\ref{eq-J_12}) is bounded in modulus by
\begin{equation*}
  q_\infty^{ (1)} \eps \int_0^{\frac{s}{\eps }} \D x\,x | \gamma (x)| \leq q_\infty^{ (1)} \eps   r \Big( \frac{s}{\eps} \Big) \;.
\end{equation*}  
Similarly, $\qtrans ( s , s- \eps x )$ can be substituted by $\qtrans (s,s)$ in the first integral over $x$  in 
(\ref{eq-J_12}), making an error  bounded by the same expression. Thus if $m\neq 1$, 
\begin{equation}\label{eq-approx_J_12}
J_{1 \rightarrow 2}^{(\lambda,\eps)} (t) = \lambda^2 \eps \re
   \int_0^t \D s  \, b_{12} (s)^2 \qtrans (s,s ) \int_0^{\frac{s}{\eps}} \D x \, \E^{(- \I  \dynphase{1}{2}+\I \zeta_{12} -\eta_{12}) (s,s-\eps x)}
     \gamma (-x)  + O (\lambda^2 \eps^{1+m_1} )
     \;.
\end{equation}     
As before, when $m=1$ the error term must be replaced by $O(\lambda^2 \eps^2 | \ln \eps|)$.

Let us introduce  an exponent  $\alpha \in (0,1/2)$. Dividing the integration range of the $s$-integral in the \RHS of (\ref{eq-approx_J_12})
into $[0, \eps^{1-\alpha}]$ and $[\eps^{1-\alpha},t  ]$ and noting that the integral over $[0, \eps^{1-\alpha}]$ can be bounded by
  $C   \eps^{1-\alpha}$ with $C=\sup_{0 \leq u \leq 1} |b_{12} (u)|^2  q_\infty^{(0)} \int_0^\infty |\gamma| < \infty$,
  one has 
\begin{eqnarray} \label{eq-approx_J_12bis}
\nonumber  J_{1 \rightarrow 2}^{(\lambda,\eps)} (t)
  & =  & \lambda^2 \eps \re 
   \int_{\eps^{1-\alpha}}^t \D s  \, b_{12} (s)^2 \qtrans (s,s ) \int_0^{\frac{s}{\eps}} \D x \, \E^{(- \I  \dynphase{1}{2}+\I \zeta_{12} -\eta_{12}) (s,s-\eps x)}
   \gamma (-x) \\
   & & + O( \lambda^2 \eps^{2-\alpha}) +   O (\lambda^2 \eps^{1+m_1} )
     \;,
\end{eqnarray}     
with the aforementioned substitution of the last error term when $m=1$.

We now divide the integration range of the $x$-integral in (\ref{eq-approx_J_12bis}) into $[0,\eps^{-\alpha}]$ and
$[\eps^{-\alpha},s/\eps]$. The integral over $[\eps^{-\alpha},s/\eps]$ is bounded
by a constant times $\eps^{\alpha m}/m$ by using the inequality $| \gamma(t)| \leq \kappa t^{-m-1}$, which follows from Asumption~\ref{assum-decay_gamma}.
Now,  for any $s \geq \eps x \geq 0$ one has
\begin{equation} \label{eq-expansion_dynphase}
  \dynphase{1}{2} (s,s-\eps x) = x e_{12}(s)  + O (\eps x^2)
\end{equation}
(see (\ref{eq-dyn_phase})). Furthermore, it follows from (\ref{eq-exact_expression_zeta}) and (\ref{eq-exact_expression_eta}) that
\begin{eqnarray} \label{eq-expansion_eta_zeta}
  \nonumber
|\zeta_{12} (s,s-\eps x) |
& =  &
  \frac{\lambda^2}{2 \eps} \bigg| \int_{s-\eps x}^s \D u \int_0^{\frac{u}{\eps}} \D y
    \big( b_1(u) b_1(u-\eps y) - b_2 (u) b_2 (u-\eps y) \big) \gamma_I (y) \bigg| 
\\ \nonumber
& \leq   &
 x \lambda^2 { \sup_{0 \leq u \leq 1, i=1,2} \{ |b_i(u)|^2\} }   \int_0^{\infty} |\gamma  |  
\\
| \eta_{12} (s,s-\eps x) |
& \leq &
 5 x \lambda^2  { \sup_{0 \leq u \leq 1, i=1,2} \{ |b_i(u)|^2\} }   \int_0^{\infty} |\gamma |\;.  
\end{eqnarray}
Taking advantage of  (\ref{eq-expansion_dynphase}) and (\ref{eq-expansion_eta_zeta}), one obtains
for any $s>\eps^{1-\alpha}$
\begin{equation*}
\int_0^{\frac{s}{\eps}} \D x \, \E^{(- \I  \dynphase{1}{2}+\I \zeta_{12} -\eta_{12}) (s,s-\eps x)}
\gamma (-x)
 = \int_0^{\infty} \D x \, \E^{- \I  x e_{12}(s)}
\gamma (-x) + O ( \eps^{1-2 \alpha}   ) + O ( \lambda^2 \eps^{-\alpha} )  + O ( \eps^{m \alpha} ) \;.
\end{equation*}
The real part of the integral in the \RHS is easily found using the symmetry properties of $\gamma$ to be equal to
half of the Fourier transform of $\gamma$ evaluated at $\omega=  e_{12}(s)$,
\begin{equation*}
\widehat{\gamma} ( e_{12}(s)) \equiv \int_{-\infty}^{\infty} \D x \, \E^{\I  x e_{12}(s)} \gamma (x) \;.
\end{equation*}
Noting that both error terms in (\ref{eq-approx_J_12}) can be dropped
  (actually,  $\eps^{m_1}\ll \max\{ \eps^{m\alpha}, \eps^{1-2\alpha}\}$ for $m \neq 1$ and
  $\eps \ln |\eps| \ll \max\{\eps^{m\alpha}, \eps^{1-2\alpha}\}$ for $m=1$),
the statement of the corollary follows.
\hfill{\finpro}

\vspace{4mm}

\Proofof{Proposition~\ref{prop-1}}
Let us denote by $\dot{b}_i(t)$ and $\ddot{b}_i(t)$ the first and second derivatives of $b_i(t)$, with 
$i= 1$, $2$ or $12$, and let us set 
\begin{equation} \label{eq-def_M_t}
 M = \sup_{0 \leq u \leq 1} \max \big\{ | b_i(u)| , |\dot{b}_i(u)| , |\ddot{b}_i (u)|\;;\;i = 1,2 \big\} \;,
\end{equation}
which is finite by assumption~\ref{assum-smoothness}. By means of a change of variables $v\to  x=(u-v)/\eps$  and the Taylor expansion
$b_i (u - \eps x) = b_i (u) - \eps x \dot{b}_i (w_{u,\eps x})$ with 
$w_{u,\eps x}$ lying between $u$ and $u - \eps x$,  
one deduces from (\ref{eq-exact_expression_eta}) that 
\begin{align*}
 & \eta_{12}(s,\tau) = \frac{\lambda^2}{2\eps} \int_\tau^s \D u \bigg[ \onehalf b_{12}(u) \int_{\frac{u-s}{\eps}}^{\frac{u-\tau}{\eps}}
    \D x  \big( b_{12} (u) -\eps x \dot{b}_{12} (w_{u,\eps x}) \big) \gamma_R (x)
    \\ \hspace{3mm}
  &    + \I  b_{12} (u) \int_{\frac{u-s}{\eps}}^{\frac{u}{\eps}} \D x \big( b_1 (u) - \eps x \dot{b}_1 (w_{u,\eps x}) \big)   \gamma_I (x) 
    - \I b_{2} (u)  \int_{\frac{u-\tau}{\eps}}^{\frac{u}{\eps}} \D x  \big( b_{12}(u) - \eps x \dot{b}_{12} (w_{u,\eps x}) \big)
     \gamma_I (x) ) \bigg]\;.
\end{align*}
It follows from (\ref{eq-vanishing_int_gamma_R}) and the symmetry properties $\gamma_R(-t)= \gamma_R(t)$, $\gamma_I (-t)= - \gamma_I (t)$ of the auto-correlation function that for $\tau \leq u \leq s$,
\begin{equation*}
  \int_{\frac{u-s}{\eps}}^{\frac{u-\tau}{\eps}} \gamma_R(x)\D x
  =   - \int_{\frac{s-u}{\eps}}^\infty \gamma_R(x)\D x  - \int_{\frac{u-\tau}{\eps}}^\infty \gamma_R(x)\D x  
\quad , \quad 
\int_{\frac{u-s}{\eps}}^{\frac{u}{\eps}} \gamma_I(x)\D x
 =   \int_{\frac{s-u}{\eps}}^{\frac{u}{\eps}} \gamma_I(x)\D x
\end{equation*}  
which implies
\begin{eqnarray*}
\bigg|  \int_{\frac{u-s}{\eps}}^{\frac{u-\tau}{\eps}} \gamma_R(x)\D x \bigg|
  & \leq &
  \int_{\frac{s-u}{\eps}}^\infty |\gamma_R(x)| \D x  + \int_{\frac{u-\tau}{\eps}}^\infty |\gamma_R (x)| \D x
  \\
 \bigg|  \int_{\frac{u-s}{\eps}}^{\frac{u}{\eps}} \gamma_I(x)\D x \bigg|
  & \leq & 
  \max \bigg\{  \int_{\frac{s-u}{\eps}}^\infty |\gamma_I(x)| \D x \, , \, \int_{\frac{u}{\eps}}^\infty |\gamma_I (x)| \D x \bigg\}\;.
\end{eqnarray*}
{ Let  $0 \leq \tau \leq s \leq t$.} Then $|\eta_{12}(s,\tau)|$ can be bounded  by 
\begin{eqnarray*}
  \nonumber
  |\eta_{12}(s,\tau)| & \leq & \frac{\lambda^2 M^2}{\eps} \int_\tau^s \D u
  \bigg( \int_{\frac{s-u}{\eps}}^\infty \big(  | \gamma_R (x)|+ |\gamma_I (x)|\big) \D x
  + \int_{\frac{u-\tau}{\eps}}^\infty \big( | \gamma_R (x)|+ | \gamma_I (x)|\big) \D x
  \\ \nonumber
  & &  
  +   \int_{\frac{u}{\eps}}^{\infty} | \gamma_I (x)|\D x   \bigg)
  \\ \nonumber
  & &
  + \lambda^2 M^2 \int_\tau^s \D u
  \bigg(  \int_{\frac{u-s}{\eps}}^{\frac{u-\tau}{\eps}} \D x\, | x \gamma_R (x)| + \int_{\frac{u-s}{\eps}}^{\frac{u}{\eps}} \D x \, |x \gamma_I (x)|
  + \int_{\frac{u-\tau}{\eps}}^{\frac{u}{\eps}} \D x\, |x \gamma_I (x)| \bigg)
  \\
  & \leq &
  \lambda^2 M^2 \bigg( 4 \int_0^{\frac{s-\tau}{\eps}} \D y \int_y^\infty | \gamma (x)| \D x
 + \int_{\frac{\tau}{\eps}}^{\frac{s}{\eps}} \D y \int_y^\infty | \gamma (x)|\D x 
  + 5 (s-\tau) \int_0^{\frac{s}{\eps}} \D x \, x |\gamma(x)| \bigg)\;, 
\end{eqnarray*}
where the last inequality is obtained by making the changes of variables $u\to y=(s-u)/\eps$,  $u\to y=(u-\tau)/\eps$,  and $u \to y=u/\eps$ and by using the parity of $|x \gamma (x)|$. { Noting that $0 \leq s-\tau \leq s \leq t \leq 1$, this gives
  $|\eta_{12}(s,\tau)| \leq 5 \lambda^2 M^2 r (t/\eps)$.

  If $0 \leq s \leq \tau \leq t$, all the estimates above remain valid provided $s$ and $\tau$ are exchanged.
  This yields the { first bound} in the Proposition.

Similarly, the derivative of $\eta_{12}(s,\tau)$ is found from (\ref{eq-exact_expression_eta}) to be given by
\begin{align} \label{eq-expression_first_derivative_eta_12} 
 & \partial_\tau \eta_{12}(s,\tau) = - \frac{\lambda^2}{2\eps}
  \bigg[ \int_{\frac{\tau-s}{\eps}}^{0} \D x  \,b_{12} (\tau ) {b}_{12} (\tau-\eps x ) \gamma_R (x)
   + \I  \int_{0}^{\frac{\tau}{\eps}} \D x \big( b_{12} (\tau) {b}_1 (\tau-\eps x) 
    \\ \nonumber
  &  \hspace*{5mm}
    - b_2 (\tau) b_{12} ( \tau-\eps x) \big)   \gamma_I (x) + \I \int_0^{\frac{s-\tau}{\eps}} \D x  \big( {\color{black}-} b_{12}(\tau) b_1 ( \tau+\eps x) + b_2 (\tau+\eps x) b_{12} (\tau) \big)
     \gamma_I (x) ) \bigg]\;,
\end{align}
which can be bounded by proceeding as above,
\begin{equation*}
| \partial_\tau \eta_{12}(s,\tau) |  \leq  \frac{2 \lambda^2 M^2}{\eps}
  \bigg[  2 \int_{\frac{|s-\tau|}{\eps}}^\infty |\gamma|  +  \int_{\frac{\tau}{\eps}}^\infty |\gamma|  
  + 2 \eps \int_0^{\frac{|s-\tau|}{\eps}} \D x \,x |\gamma(x) | +  \eps \int_0^{\frac{\tau}{\eps}} \D x \, x |\gamma (x) | \bigg] \;.
\end{equation*}
Integrating this expression with respect to $\tau$ { from $0$ to $t$} and making the changes of variables $\tau\to y=|s-\tau|/\eps$
and $\tau \to y=\tau/\eps$, the { second estimate} in the Proposition is obtained.
Using (\ref{eq-expression_first_derivative_eta_12}) again, one easily proves that 
\begin{equation*}
\sup_{\tau , s \in [0,t]} | \partial_\tau \eta_{12}(s,\tau) |  \leq  \frac{6 \lambda^2 M^2}{\eps} \int_0^\infty | \gamma|
\;.
\end{equation*}
This inequality together { with the first bound  yields the third} bound in the Proposition.

The second derivative of $\eta_{12}(s,\tau)$ is found to be equal to
%
\begin{eqnarray} \label{eq-second_derivative_eta}
\nonumber  
  \partial_\tau^2 \eta_{12}(s,\tau)
  & = & \frac{\lambda^2}{2\eps^2} \bigg( b_{12}(\tau) b_{12} (s) \gamma \Big( \frac{\tau-s}{\eps} \Big)
  - \I \big( b_{12}(\tau) b_1 (0) - b_2 (\tau) b_{12}(0) \big)  \gamma_I \Big( \frac{\tau}{\eps} \Big) \bigg)
  \\ \nonumber
  & & - \frac{\lambda^2}{2 \eps} \partial_\tau (b_{12}^2 ) (\tau) \bigg( \int_{ \frac{\tau-s}{\eps}}^0 \gamma_R
  - \I \int_{- \frac{\tau}{\eps}}^{ \frac{s- \tau}{\eps}} \gamma_I \bigg)
  \\
  & & + O \bigg( \lambda^2 \int_0^{\frac{s}{\eps}} \D x \,x|\gamma(x)| \bigg) \;.
\end{eqnarray}
In this expression, we have approximated $b_i(\tau - \eps x)$ and $\dot{b}_i(\tau - \eps x)$ by $b_i(\tau)$ and
$\dot{b}_i (\tau)$, for $i=1,2$, and  $12$. The error incurred on $\partial_\tau^2 \eta_{12}(s,\tau)$ is estimated, using 
$|b_i (\tau - \eps x)- b_i (\tau)| \leq M \eps |x|$ and
$|\dot{b}_i(\tau - \eps x)- \dot{b}_i (\tau)| \leq M \eps |x|$ with 
$M$ given by (\ref{eq-def_M_t}), to be less than $12 M^2 \lambda^2 \int_0^{\frac{s}{\eps}} \D x \,x|\gamma(x)|$.

Let us set
\begin{equation*}
  I(s) \equiv \int_0^s \D \tau \,h(s,\tau) \partial_\tau^2 \eta_{12} (s,\tau)\;.
\end{equation*}
The contribution of the first line in (\ref{eq-second_derivative_eta}) to this integral is
given after making the changes of variables $x=(s-\tau)/\eps$ and $x=\tau/\eps$ by 
\begin{equation*}
  \frac{\lambda^2}{2 \eps} \int_0^{\frac{s}{\eps}} \D x \bigg( h(s,s-\eps x) b_{12}(s-\eps x) b_{12}(s) \gamma (-x)
  -\I h(s,\eps x ) \big( b_{12}(\eps x) b_1(0) - b_2(\eps x) b_{12}(0) \big) \gamma_I (x) \bigg)\;.
\end{equation*}  
In this expression, $ b_i(s-\eps x)$ and  $b_i(\eps x)$ can be replaced by $b_i(s)$ and $b_i(0)$, respectively,
making an error $\Oo(\lambda^2 \int_0^{s/\eps} \D x \,x |\gamma(x)|)$. 
The contribution to the integral  $I(s)$ of the second line  in (\ref{eq-second_derivative_eta}) is bounded by
$8 N M^2 \lambda^2 \int_0^{s/\eps} \D y \int_y^\infty | \gamma|$, making use once again of (\ref{eq-vanishing_int_gamma_R}) and the symmetry
properties of $\gamma_R$ and $\gamma_I$.
 The error term of order $\lambda^2 \int_0^{\frac{s}{\eps}} \D x \,x|\gamma(x)|$ produces an error of the same 
order in $I(s)$.
This proves the last bound in the Proposition.
\hfill{\finpro}

\subsubsection{Estimations on the derivatives of $\zeta_{12}$}
\label{sec_estimations_zeta}

We show in this subsection that $\zeta_{12}(s,\tau)$ is of order $\lambda^2/\eps$. More precisely, it behaves as
\begin{equation} \label{eq-behavior_zeta_12}
  \zeta_{12}(s,\tau) =  -\frac{\lambda^2 \beta_0}{2 \eps} \int_\tau^s \D u \, ( b_1^2 - b_2^2 )( u) + {O}(\lambda^2 \eps^{m-1})\; 
\end{equation}
 as $(\eps,\lambda) \to (0,0)$, with
  \begin{equation}
    \beta_0 \equiv \int_0^\infty \D x \,\gamma_I (x)\;.
  \end{equation}  
Furthermore, we prove that
$\partial_\tau^2 \zeta_{12}(s,\tau) $ can be approximated
by the
 second derivative of the first term in the \RHS of (\ref{eq-behavior_zeta_12})  in certain integrals
$\int_0^s \D \tau \,h(s,\tau) \partial^2_\tau \zeta_{12} (s,\tau)$,  if
$h(s,\tau)$ is a smooth enough function vanishing for $\tau=0$.
{ Note that $\partial_\tau^n \zeta_{12} (s,\tau) = - \partial_\tau^n \zeta_{12}(\tau)$ for $n=1,2,\ldots$, since by definition
  $\zeta_{12} (s,\tau) = \zeta_{12}(s)-\zeta_{12}(\tau)$.}

\begin{prop} \label{prop-5}
  Suppose that $\gamma \in L^1 (\real)$  and that assumption~\ref{assum-smoothness} holds.
   Let $h_{\eps,\lambda} (s,\tau)\equiv q (s,\tau) g_{\eps,\lambda}(s,\tau)$ be such  $q, g_{\eps,\lambda} \in C^1([0,1]^2)$,
  $q$ is independent of $(\eps,\lambda)$, and 
  \begin{equation*}
   q(s,0)=0\;,\; 
   \sup_{0 \leq \tau \leq s \leq 1} | q (s,\tau) |  < \infty \;,\;
   \sup_{0 \leq \tau \leq s \leq 1} | \partial_\tau q (s,\tau) | < \infty
   \;\text{ and } \; \sup_{0 \leq \tau \leq s \leq 1} | g_{\eps,\lambda} (s,\tau) | \leq 1 \;. 
  \end{equation*}
  Then there is a constant $c'$ independent on $\eps,\lambda$ such that
  for any $ 0 \leq t \leq 1$,
  \begin{equation}
    \sup_{\tau \in [0,t]}  | \partial_\tau \zeta_{12} (\tau) | \leq c' \frac{\lambda^2}{\eps}   \quad
    , \quad   \int_0^{ t} \D \tau\, | \partial_\tau^2 \zeta_{12} (\tau) |
    \leq  c' \frac{\lambda^2}{\eps}  
  \end{equation}
  and for any $ 0 \leq \tau \leq s \leq 1$,
  \begin{equation} \label{eq-bound_on_zeta_and_derivatives2}
  \begin{array}{l} \displaystyle
  \Big| \zeta_{12}(s,\tau) + \frac{\lambda^2 \beta_0}{2 \eps} \int_\tau^s \D u \, ( b_1^2 - b_2^2 )( u) \Big|
  \quad  \leq \quad  c' \lambda^2  r \Big( \frac{s}{\eps}  \Big) 
  \\[4mm] \displaystyle
   \bigg|
     \int_0^s \D \tau \, h_{\eps,\lambda} (s,\tau) \partial^2_\tau \zeta_{12} (s,\tau)
     - \frac{\lambda^2  \beta_0 }{2 \eps} \int_0^{s} \D \tau \, h_{\eps,\lambda} (s,\tau) \partial_\tau ( b_1^2 - b_2^2)(\tau)
   \bigg| \quad  \leq \quad  c' \lambda^2 r  \Big( \frac{s}{\eps}  \Big) 
  \end{array}
 \end{equation}
  with $r(z)$ defined in (\ref{eq-def_d(z)}). 
 \end{prop}

In analogy with the developments above,
let us consider the term involving $\partial_\tau^2 \zeta_{12}(s,\tau)$ in
 the formula  (\ref{eq-IPP_formula_first_term}) for the transition probability,
\begin{equation} \label{eq-def_L_12}
L_{1 \rightarrow 2}^{(\lambda,\eps)} (t)  \equiv  
    - 2 \eps^2 \re \left\{ \I \int_0^t \D s \int_0^s \D \tau\,  \E^{ (- \I \dynphase{1}{2}+\I \zeta_{12} -\eta_{12}) (s,\tau)} \partial_\tau^2 \zeta_{12} (s,\tau) \qtrans (s,\tau)
    \right\}\;.
\end{equation}
%

\begin{cor} \label{corr-corollary-estimation_L_12}
  Assume~{\ref{assum-gap}-\ref{assum-decay_gamma}}. Then  
  \begin{eqnarray}
    \nonumber
    L_{1 \rightarrow 2}^{(\lambda,\eps)} (t) & = &
     -  \lambda^2 \eps \beta_0
    \re \bigg\{ \I \int_0^t \D s \int_0^s \D \tau \, \E^{ (- \I \dynphase{1}{2}+\I \zeta_{12} -\eta_{12}) (s,\tau)}  \partial_\tau ( b_1^2 - b_2^2) (\tau) \qtrans (s,\tau) \bigg\}
    \\
    & & + \Oo(\lambda^2 \eps^{1+m_{1}} |\ln \eps |^{\delta_{m,1}} ) 
  \end{eqnarray}
  with $\delta_{m,1}=1$ if $m=1$ and $0$ otherwise.
\end{cor}  

\Proofof{Corollary~\ref{corr-corollary-estimation_L_12}}
One applies Proposition~\ref{prop-5} with
$g_{\eps,\lambda}=\E^{- \I \dynphase{1}{2}+\I \zeta_{12} -\eta_{12} }$ and $q= \qtrans$, which satisfy the required assumptions
since $\qtrans (s,0)=0$ and $\re \eta_{12} \geq 0$, see also (\ref{eq-uniform_bound_q}). The statement
then follows from (\ref{eq-bound_on_r}).
\hfill{\finpro}

\vspace{4mm}

\Proofof{Proposition~\ref{prop-5}}
The derivatives of $\zeta_{12}(s,\tau)$ are equal to (see (\ref{eq-exact_expression_zeta}))
\begin{equation} \label{eq-exact_expression_derivative_zeta}
\begin{array}{lclcl} 
  \partial_\tau \zeta_{12} (s,\tau)
  & = & { - \partial_\tau \zeta_{12} (\tau)} & = &  
 \dss  \frac{\lambda^2}{2 \eps} \int_0^{\frac{\tau}{\eps}} \D x\, \Big( b_1(\tau ) b_1(\tau -\eps x) -b_2(\tau ) b_2(\tau -\eps x)\Big) \gamma_I(x)
  \\[4mm]
  \partial^2_\tau \zeta_{12} (s,\tau)
  & = & { - \partial^2_\tau \zeta_{12}(\tau)} &  = &  
 \dss  \frac{\lambda^2}{2\eps}  \int_0^{\frac{\tau}{\eps}} \D x\, \partial_\tau \big( b_1(\tau) b_1(\tau -\eps x) -b_2(\tau) b_2(\tau -\eps x)\big) \gamma_I(x)
  \\[4mm]
  & & & &  \dss
   + \frac{\lambda^2}{2 \eps^2} \big( b_1(\tau) b_1(0)-b_2(\tau) b_2 (0) \big) \gamma_I \Big( \frac{\tau}{\eps} \Big) \;.
\end{array}
\end{equation}
Clearly, for any $0 \leq \tau \leq t \leq 1$,
$| \partial_\tau \zeta_{12}(\tau) |$ and $\int_0^t \D \tau\,| \partial^2_\tau \zeta_{12} (\tau) |$ are bounded from above 
 by $c' (\lambda^2/\eps)$ and by $3 c'(\lambda^2/\eps)$, respectively, with $c' =M^2 \int_{0}^\infty |\gamma_I | < \infty$,} $M$ being given by (\ref{eq-def_M_t}).
The two last bounds in the Proposition follow from (\ref{eq-exact_expression_zeta}) and (\ref{eq-exact_expression_derivative_zeta}), by relying on similar arguments as those of the proof of Proposition~\ref{prop-1} and making use of the Taylor expansion $q(s,\eps x) = \eps x \partial_\tau q ( s , w_{s,\eps x} )$ with $w_{s,\eps x} \in [0, \eps x]$.

\hfill{\finpro}

\subsection{Adiabatic limit of the first term of the Dyson series} \label{sec-behavior_first_term_Dyson}

We are now ready to estimate the \RHS of the integration by part formula  (\ref{eq-IPP_formula_first_term}).
The contributions of the different terms generated by the derivatives with respect to $\tau$ 
can be estimated by relying on Propositions~\ref{prop-1} and~\ref{prop-5}. The strategy  is as follows:
\begin{enumerate}[label=(\Roman*)]
\item  \label{it-strategy_i} show with the help of Corollary~\ref{eq-corollary-bound_int_derivative_q} that   the terms involving the first derivative of $\eta_{12} (s,\tau)$ contribute to order $\Oo( \lambda^2 \eps^{1+m_1}  | \ln \eps |^{\delta_{m,1}} ) + \Oo(\lambda^4 \eps^{m_{1}}  | \ln \eps |^{\delta_{m,1}} ) $;
\item \label{it-strategy_ii} show by means of another integration by parts
  that the terms involving derivatives of $e_{12}$ and $\qtrans$  and the first derivative of $\zeta_{12}$ only
  contribute to order $\Oo(\eps^3) + \Oo(\lambda^2 \eps^2)+\Oo(\lambda^4 \eps)+\Oo(\lambda^6)$;
\item \label{it-strategy_iii} show with the help of Corollary~\ref{corr-corollary-estimation_J_12} that   the terms involving the second derivative of $\eta_{12} (s,\tau)$ can be approximated by
  the second term in the \RHS of (\ref{eq-main_result}), which is of order $\lambda^2 \eps$;
\item   \label{it-strategy_iv} show with the help of Corollary~\ref{corr-corollary-estimation_L_12} and an integration by parts
  that   the terms involving the second derivative of $\zeta_{12} (s,\tau)$ contribute to order
  ${\Oo(\lambda^2 \eps^{1+m_1} | \ln \eps |^{\delta_{m,1}} )} + \Oo(\lambda^4 \eps )$.
\end{enumerate}
Note that the error terms obtained in \ref{it-strategy_i} are small with respect to both contributions to the transition probability in (\ref{eq-main_result}) (that is, to the transition probability in the absence of reservoir and to our estimated correction coming from the coupling to the reservoir)
provided that { $\lambda^2 \ll \eps^{1-\frac{m_1}{2}} \ll 1$ if $m\not= 1$ and $\lambda^2 \ll ( \eps/|\ln \eps| )^{1/2}$ if $m=1$.}
Similarly, the  error terms obtained in \ref{it-strategy_ii} are negligible with respect to both contributions
provided that $\eps^2 \ll \lambda^2 \ll \eps^{2/3} \ll 1$.

We implement our strategy by first expressing
$ \| \omeg{1} (t) \|^2$ as a sum of six terms. { From (\ref{eq-IPP_formula_first_term}),}
\begin{equation} \label{eq-decomp_norm_omeg1}
  \| \omeg{1} (t) \|^2  = \eps^2 q_{1\rightarrow 2}(t,t)  + I_{1 \to 2}^{(\lambda,\eps)} (t) + J_{1 \to 2}^{(\lambda,\eps)} (t) + L_{1\to2}^{(\lambda,\eps)} (t)
  + R_{1 \to 2}^{(\lambda,\eps)} (t) + S_{1 \to 2}^{(\lambda,\eps)} (t)
\end{equation}
with
\begin{eqnarray}
  \nonumber
 I_{1 \to 2}^{(\lambda,\eps)} (t) & = &
  -2 \eps^2 \re \int_0^t \D s \int_0^s \D \tau\, \Big( \E^{- \I  \dynphase{1}{2} +\I \zeta_{12} - \eta_{12}}
   \big(  \partial_\tau v_{12} + \I  ( \partial_\tau \zeta_{12} ) w_{12} 
   - ( \partial_\tau \zeta_{12} )^2 \qtrans \big)
   \Big) (s,\tau) 
    \\ \nonumber
 R_{1 \to 2}^{(\lambda,\eps)} (t) & = &
 2 \eps^2 \re \int_0^t \D s \int_0^s \D \tau\, \Big( \E^{- \I  \dynphase{1}{2} +\I \zeta_{12} - \eta_{12}}
 (\partial_\tau \eta_{12})  w_{12} \Big) (s,\tau)
 \\ \nonumber
 S_{1 \to 2}^{(\lambda,\eps)} (t) & = &
 -2 \eps^2 \re \int_0^t \D s \int_0^s \D \tau\, \Big( \E^{- \I  \dynphase{1}{2} +\I \zeta_{12} - \eta_{12}}
 \Big( - 2 \I ( \partial_\tau \zeta_{12} )( \partial_\tau \eta_{12}) + (\partial \eta_{12})^2 \Big) \qtrans \Big)(s,\tau)
\end{eqnarray}  
and $J_{1 \to 2}^{(\lambda,\eps)} (t)$ and $L_{1 \to 2}^{(\lambda,\eps)} (t)$ are
defined in (\ref{eq-def_J_12}) and (\ref{eq-def_L_12}).
Here, we have set 
\begin{equation*}
  v_{12}  =  \partial_\tau ( \ln |e_{21}| ) \qtrans + \partial_\tau \qtrans
  \qquad , \qquad w_{12}  =\partial_\tau ( \ln |e_{21}|) \qtrans + 2 \partial_\tau \qtrans\;.
\end{equation*}
 We assume in the sequel that Assumptions~\ref{assum-gap}-\ref{assum-decay_gamma} are satisfied. Then
\begin{equation} \label{eq-boundedness_v_w}
\sup_{0 \leq \tau \leq  1} \max_{n=1,2,3} | \partial_\tau^n ( \ln |e_{21}|) (\tau) | < \infty
\quad , \quad  a \equiv 
\sup_{0 \leq \tau \leq s \leq   1} \max_{n=0,1} \{ | \partial_\tau^{n+1} v_{12}(s,\tau) | , | \partial_\tau^{n} w_{12} (s,\tau)| \}  < \infty \;.
\end{equation}
Actually, one can bound e.g. $| \partial_\tau ( \ln | e_{12}| ) (\tau) |$ by $2 \max_{i=1,2} \sup_{0 \leq \tau \leq 1}  |\partial_\tau e_i(\tau)| /\delta$;
 the boundedness of the second supremum is a consequence of that of the first one and of (\ref{eq-uniform_bound_q}).

We now proceed to prove the statements \ref{it-strategy_i}-\ref{it-strategy_iv}.

\begin{itemize}
\item[\ref{it-strategy_i}]  
 Using (\ref{eq-boundedness_v_w}), applying Corollary~\ref{eq-corollary-bound_int_derivative_q} and recalling that $\re \eta_{12} (s,\tau) \geq 0$, one has
\begin{equation*}
  |  R_{1 \to 2}^{(\lambda,\eps)} (t) | \leq 2 a c  \lambda^2  \eps^{1+m_1} |\ln \eps|^{\delta_{m,1}}
\;. \end{equation*}    
Similarly, using the first bound in Proposition~\ref{prop-5} and the same Corollary,  
\begin{equation*}
  |  S_{1 \to 2}^{(\lambda,\eps)} (t) | \leq 2 ( c c'  + c ) q_\infty \lambda^4 \eps^{m_1}  |\ln\eps |^{\delta_{m,1}} \;.
\end{equation*}
Hence $ R_{1 \to 2}^{(\lambda,\eps)} (t)$ and $ S_{1 \to 2}^{(\lambda,\eps)} (t)$ are of order $\lambda^2 \eps^{1+m_1} |\ln\eps |^{\delta_{m,1}}$
and $\lambda^4 \eps^{m_1} |\ln\eps |^{\delta_{m,1}}$, respectively,
as annouced above.

\item[\ref{it-strategy_ii}] 
We now estimate   $I_{1 \to 2}^{(\lambda,\eps)} (t)$ by performing an integration by parts  in the $\tau$-integral,
using $e_{21} (\tau) \E^{-\I \dynphase{1}{2}(s,\tau)}= \I \eps  \partial_\tau ( \E^{-\I \dynphase{1}{2} (s,\tau)})$.
By relying on observation~\ref{item-1_prof_lemma2} in the proof of Proposition
\ref{lem-2},  one has  $\partial_\tau^n \qtrans (s,0)=0$ for $n=0,1,2$, showing that
$\partial_\tau v_{1\to 2}(s,0)=w_{1\to 2}(s,0)=0$.
Using also
the fact that $\qtrans (s,s)$ and $\partial_\tau \zeta_{12} (s)$ are real
and the observation~\ref{item-2_prof_lemma2} of the same proof, one finds that the boundary term in the integration by parts reads
\begin{equation*}
  2 \eps^3 \int_0^t \D s \,
    \frac{1}{e_{21}(s)} \Big(  -\I \partial_\tau v_{12} (s,s) + ( \partial_\tau \zeta_{12} ) ( s)  w_{12} (s,s) \Big)\;.
\end{equation*}
One then deduces from (\ref{eq-boundedness_v_w}),  the gap hypothesis~\ref{assum-gap}, and Proposition~\ref{prop-5} that the  boundary term
is of order $\Oo(\eps^3)+\Oo(\lambda^2 \eps^2)$.
Thus the integration by parts yields
\begin{eqnarray} \label{eq-proof_Prop4.7II}
  \nonumber
  I_{1 \to 2}^{(\lambda,\eps)} (t)
& = & 2
 \eps^3 \re
 \int_0^t \D s \int_0^s \D \tau\,  \E^{- \I  \dynphase{1}{2} (s,\tau)} \partial_\tau
 \Bigg[ \frac{\E^{\I \zeta_{12} - \eta_{12}}}{e_{21}(\tau)} 
   \big(  \I \partial_\tau v_{12} -  ( \partial_\tau \zeta_{12} ) w_{12}  
   \\
& &    - \I ( \partial_\tau \zeta_{12} )^2 \qtrans \big)
   \bigg] (s,\tau) + \Oo(\eps^3)+\Oo(\lambda^2 \eps^2) \;.
\end{eqnarray}
Computing the derivative of the expression inside the square brackets and using
(\ref{eq-uniform_bound_q}), (\ref{eq-boundedness_v_w}) and the gap hypothesis~\ref{assum-gap}, one finds that
 for any $0 \leq \tau \leq 1$, the
integrand in (\ref{eq-proof_Prop4.7II}) is bounded in modulus by a constant factor  times the sum of terms
\begin{equation*}
\sum_{n=0}^3 | \partial_\tau \zeta_{12} (\tau)|^n 
 + | \partial_\tau^2 \zeta_{12} (\tau)|  
+ | \partial_\tau^2 \zeta_{12} (\tau)|  | \partial_\tau \zeta_{12} (\tau)| +
| \partial_\tau \eta_{12} (s,\tau) | \sum_{n=0}^2 | \partial_\tau \zeta_{12} (\tau)|^{n}.
\end{equation*}

Thanks to Corollary~\ref{eq-corollary-bound_int_derivative_q} and Proposition~\ref{prop-5}, the double integrals
 of this sum in the triangle $\{ 0 \leq \tau \leq s \leq t\}$ is bounded from above  by
\begin{equation*}
\frac{t^2}{2} \sum_{n=0}^3 (c')^n  \frac{\lambda^{2n}}{\eps^n} +
t c' \frac{\lambda^2}{\eps} +
t (c')^2 \frac{\lambda^4}{\eps^2} +
 \lambda^2 \eps^{m_1-1}  |\ln \eps |^{\delta_{m,1}} \sum_{n=0}^2 (c')^{n} \frac{\lambda^{2n}}{\eps^n}  \;.
\end{equation*}
This proves that
\begin{equation} \label{eq-LO_I_12}
I_{1 \to 2}^{(\lambda,\eps)} (t)
= \Oo( \eps^3) + \Oo( \lambda^2 \eps^2 ) + \Oo(\lambda^4 \eps) + \Oo(\lambda^6)\;,
\end{equation}
as annouced above.

\item[\ref{it-strategy_iii}]
  By virtue of Corollary~\ref{corr-corollary-estimation_J_12},
  \begin{equation} \label{eq-proof_Prop4.7III}
   J_{1 \rightarrow 2}^{(\lambda,\eps)} (t) = \frac{\lambda^2 \eps}{2}  \int_0^t \D s\, b_{12}^2(s) \qtrans (s,s) \widehat{\gamma} ( e_{12}(s) ) + \Oo( \lambda^2 \eps^{\min\{1+ m\alpha ,2 - 2 \alpha\}} ) + \Oo( \lambda^4 \eps^{1-\alpha} ) 
  \end{equation}
  with $\alpha \in (0,1/2)$ an arbitary exponent.
  We choose $\alpha$ as follows. First, since we want that the aforementioned errors to be much smaller than  
  both contributions to the transition probability in (\ref{eq-main_result}), we require that in the limit
  $\eps \ll 1$,
  \begin{equation*}
  \left\{ \begin{array}{lcl}
    \lambda^2 \eps^{1+m\alpha} & \ll & \min \{ \eps^2\, ,\,\lambda^2 \eps \}  \\
    \lambda^2 \eps^{2-2\alpha} & \ll & \min \{ \eps^2\,,\,\lambda^2 \eps \}\\
    \lambda^4 \eps^{1-\alpha} & \ll &  \min \{ \eps^2\,,\,\lambda^2 \eps \}
  \end{array} \right.
  \quad \Leftrightarrow \quad 
  \left\{ \begin{array}{lcl}
    \lambda^2  & \ll & \eps^{1-m\alpha} \\
    \lambda^2  & \ll & \eps^{2\alpha} \\
    \lambda^2  & \ll & \min \{ \eps^{\alpha}\, ,\, \eps^{(1+\alpha)/2} \}\;.
  \end{array} \right.
  \end{equation*}
   The optimal value $\alpha_0$  minimizes the maximal exponent  $\max \{ 1-m\alpha , 2 \alpha, (1+\alpha)/2\}$ of $\eps$ in the bounds on $\lambda^2$.
  One easily finds
  \begin{equation} \label{eq-alpha_0}
     \alpha_0 = \frac{1}{2+2 m - m_1} = 
     \begin{cases} \dss
      \frac{1}{1+2m} & \text{ if $m\geq 1$}
      \\[3mm] \dss
      \frac{1}{2+m}  & \text{ if $0<m\leq 1$.}
      \end{cases}
  \end{equation}
  For the choice $\alpha = \alpha_0$,   the error terms in the estimation  (\ref{eq-proof_Prop4.7III})
  are $\Oo(\lambda^2 \eps^{1+m \alpha_0})$.

In fact, if $m > 1$ then { $\lambda^2 \eps^{2-2\alpha_0} \ll \lambda^2 \eps^{1+m \alpha_0}$}
(since  $1+ m\alpha_0 < 2 - 2 \alpha_0$)  and 
  $\lambda^4 \eps^{1-\alpha_0} \ll \lambda^2 \eps^{1+m \alpha_0}$ (since by construction
$\lambda^2  \ll  \eps^{1-m\alpha_0}$  and $1-m\alpha_0= \alpha_0+m\alpha_0$). Similarly, if $0<m \leq 1$ then
 $\lambda^2 \eps^{2-2\alpha_0} = \lambda^2 \eps^{1+m \alpha_0}$
(since $1+ m\alpha_0 = 2 - 2 \alpha_0$)
and $\lambda^4 \eps^{1-\alpha_0} \ll \lambda^2 \eps^{1+m \alpha_0}$ (since by construction
$\lambda^2 \ll \eps^{(1+\alpha_0)/2}$ and $(1+\alpha_0)/2 > \alpha_0+m\alpha_0$). 
  Thus in all cases
  \begin{equation} \label{eq-LO_J_12}
    J_{1 \rightarrow 2}^{(\lambda,\eps)} (t) =  \frac{\lambda^2 \eps}{2}  \int_0^t \D s\, b_{12}^2(s) \qtrans (s,s) \widehat{\gamma} ( e_{12}(s) )
    + \Oo( \lambda^2 \eps^{1+m \alpha_0} ) \;.
  \end{equation}

\item[\ref{it-strategy_iv}]  It remains to estimate $L_{1 \to 2}^{(\lambda,\eps)} (t)$. For this, we use Corollary~\ref{corr-corollary-estimation_L_12} and make an integration by parts to get
  \begin{eqnarray} \label{eq-proof_Prop4.7IV}
  \nonumber
  L_{1 \rightarrow 2}^{(\lambda,\eps)} (t)  & = &  
   \beta_0 \lambda^2 \eps^2 \bigg\{ 
   \int_0^t \D s\,\partial_s ( b_1^2 - b_2^2 )(s) \frac{\qtrans (s,s)}{e_{21}(s)}
   \\ \nonumber
   &&
   - \re \int_0^t \D s \int_0^s \D \tau\,\E^{- \I  \dynphase{1}{2} (s,\tau)} \partial_\tau
   \bigg( \E^{(\I \zeta_{12} - \eta_{12})(s,\tau)} \partial_\tau ( b_1^2 - b_2^2 ) (\tau) \frac{\qtrans (s,\tau)}{e_{21}(\tau)} \bigg) \bigg\}
   \\
   & & + \Oo( \lambda^2 \eps^{1+m_1 }  |\ln \eps |^{\delta_{m,1}}  ) \; ,
  \end{eqnarray}
  where we have used again $\qtrans (s,0)=\dynphase{1}{2} (s, s)= \zeta_{12} (s,s)= \eta_{12} (s,s)=0$  as well as $\qtrans (s,s )\in \real$.
  The boundary term in the first line of (\ref{eq-proof_Prop4.7IV}) is obviously of order $\lambda^2 \eps^2$. 
  The integral in the second line, in turn, is bounded by
  \begin{equation*}
   \frac{4 M^2}{\delta}  \int_0^t \D s \int_0^s \D t\, \bigg( \Big(  \sup_{0 \leq \tau \leq 1} | \partial_\tau (\ln |e_{21}|)(\tau) |
    + \sup_{0 \leq \tau \leq 1} | \partial_\tau \zeta_{12} (\tau) | +  | \partial_\tau \eta_{12} (s,\tau)| +2 \Big)  q_\infty 
    + { q_\infty^{(1)}} \bigg)\;. 
  \end{equation*}
   In view of Corollary~\ref{eq-corollary-bound_int_derivative_q} and Proposition~\ref{prop-5},
  the contribution  to  $L_{1 \rightarrow 2}^{(\lambda,\eps)} (t)$ of the integral in the second line of (\ref{eq-proof_Prop4.7IV})  is thus of
    order $ \Oo(\lambda^2 \eps^2) + \Oo(\lambda^4 \eps )$. 
   This gives
 \begin{equation} \label{eq-LO_L_12}
   L_{1 \rightarrow 2}^{(\lambda,\eps)} (t) =   \Oo( \lambda^2 \eps^{1+m_1 } |\ln \eps |^{\delta_{m,1}}  ) + \Oo( \lambda^4 \eps )\;.
 \end{equation}

\end{itemize}

Altogether, {collecting (\ref{eq-decomp_norm_omeg1}), (\ref{eq-LO_I_12}), (\ref{eq-LO_J_12}) and (\ref{eq-LO_L_12}) and using
  $\lambda^2 \eps^{1+m_1 } |\ln \eps |^{\delta_{m,1}}  \ll \lambda^2 \eps^{1 + m \alpha_0}$
    (since $m \alpha_0 < m_1$)}, we deduce that 
\begin{align} \label{eq-proof_prop4.7}
  \nonumber
  \| \omeg{1} (t) \|^2 &=
 \eps^2 \qtrans (t,t)    + \frac{\lambda^2 \eps }{2} \int_0^t \D s \, \qtrans (s,s) b_{12}^2(s) \widehat{\gamma} \big( e_{12} (s) \big) \\
 &+  \Oo(\eps^3) + \Oo( \lambda^2 \eps^{1+m \alpha_0 } ) + \Oo( \lambda^4 \eps^{m_1}  | \ln \eps |^{\delta_{m,1}} ) + \Oo(\lambda^6 )\;.
  \end{align}
 Since  $ \transprobafree (t)=\eps^2 \qtrans (t,t)+ \Oo(\eps^3) $ (see (\ref{eq-transition_proba_free})),
one may substitute $\transprobafree (t)$ for
 $ \eps^2 \qtrans (t,t)$ in (\ref{eq-proof_prop4.7}),
making an error of order $\lambda^2 \eps^2 \ll \lambda^2 \eps^{1+m \alpha_0}$.
We conclude that

\vspace{1mm}

\begin{prop} \label{prop-6}
 Under assumptions \ref{assum-gap}-\ref{assum-decay_gamma}, 
  \begin{eqnarray*} \label{eq-estim_first_term_Dyson_series}
  \| \omeg{1} (t) \|^2  \nonumber
    & = &
  \transprobafree (t)    + \frac{\lambda^2}{2\eps} \int_0^t \D s \,\transprobafree ( s) b_{12}^2(s) \widehat{\gamma} \big( e_{12} (s) \big) 
  \\
  & &  +  \Oo(\eps^3) + \Oo( \lambda^2 \eps^{1+m \alpha_0 } ) + \Oo( \lambda^4 \eps^{m_1}  | \ln \eps |^{\delta_{m,1}} ) + \Oo(\lambda^6 )\;,
  \end{eqnarray*}
  where $\alpha_0$ is given by (\ref{eq-alpha_0}).
\end{prop}  


Observe that the exponent of $\eps$ in the  second error term, $1+m\alpha_0$, belongs to $(1, 4/3]$ when $0<m \leq 1$ and to $(4/3,3/2)$ when $m > 1$.

\section{Contribution of higher-order terms in the Dyson expansion} \label{sec-higher_order_terms}

Recall that the transition probability between distinct energy levels of the system is given by
\begin{equation*}
  \transproba (t)= \bigg\| \sum_{k \geq 1}  P_2 (0)  \omeg{k} (t)\bigg\|^2
  = \bigg\| \sum_{k\text{ odd}}^\infty \omeg{k} (t)\bigg\|^2\;,
\end{equation*}
see (\ref{eq-trans_proba_infinite_sum}).
In this section, we show that the terms of this series of order $k >1$
do not contribute to the transition probability  to lowest order in $\eps$ and $\lambda$. 
The main result is summarized in the following Proposition.

\vspace{2mm}

\begin{prop} \label{prop_contibution_higher_order_terms}
  Under assumptions~\ref{assum-gap}-\ref{assum-decay_gamma},  we have  
\begin{equation}  
\sum_{j=1}^\infty \sup_{0 \leq t \leq 1} \big\| \omeg{2j+1} (t) \big\|  
= \Oo\big( \eps^2 +  \lambda \eps^{\frac{1+m_1}{2}} | \ln \eps |^{\onehalf \delta_{m,1}} 
+   \lambda^2 \eps^{\frac{m_1}{2}} | \ln \eps |^{\onehalf \delta_{m,1}} +\lambda^5 \big)\;,
\end{equation}
where as above $m_1 = \min\{m,1\}$.
\end{prop}

In view of (\ref{lead}) and since
\[ 
\| \omeg{1} (t) \|= \Oo(\eps+\lambda \eps^{\onehalf } +\lambda^2\eps^{\onehalf m_1} | \ln \eps |^{\onehalf \delta_{m,1}} +\lambda^3 )
\] 
as  shown in the previous section,
one deduces from Proposition~\ref{prop_contibution_higher_order_terms} that 
\begin{eqnarray} \label{eq-trans_proba_vs_omega1}
\nonumber  
\transproba (t)
& = &
\| \omeg{1} (t) \|^2  
+ \Oo ( \eps^3  +\lambda \eps^{\frac{3+m_1}{2}} | \ln \eps |^{\onehalf \delta_{m,1}} 
+   \lambda^2 \eps^{1+\frac{m_1}{2}} | \ln \eps |^{\onehalf \delta_{m,1}}  + \lambda^3 \eps^{\frac{1+m_1}{2}} | \ln \eps |^{\onehalf \delta_{m,1}} \big)
\\
& & 
+ \Oo \big( \lambda^4 \eps^{m_1} | \ln \eps |^{\delta_{m,1}} 
 +  \lambda^5 \eps^{\onehalf m_1} | \ln \eps |^{\onehalf \delta_{m,1}} +\lambda^8 \big) 
\;.
\end{eqnarray}
Together with Proposition~\ref{prop-6}, this yields the result stated in Theorem~\ref{thm-1}.
Actually, given that $m \alpha_0 < m_1/2$, the error term proportional to $\lambda^2$ in (\ref{eq-trans_proba_vs_omega1}) is much smaller than
$\lambda^2 \eps^{1+m\alpha_0}$.

\vspace{2mm}

To prove Proposition~\ref{prop_contibution_higher_order_terms}, we proceed analogously as in the previous section. We first
integrate by parts the recursion relation (\ref{eq-iterative_formula_omega_k}) and then rely on the
estimations  of Subsection~\ref{sec-estim_eta_zeta} to bound
$\sup_{0 \leq t \leq 1} \| \omeg{2j+1} (t)\|^2$ in terms of its value for $j \rightarrow j-1$
up to some remainder terms (Subsection~\ref{eq-IPP_higher_order_terms}). 
The remainder terms involve multiple integrals of first and second derivatives of quantum expectations in the vacuum
state of products of $2j+2$ Weyl operators. The latter are
controled in Subsections~\ref{sec-estimation_D} and~\ref{sec-estimation_E} by
using similar arguments as in  Subsection~\ref{sec-estim_eta_zeta}.
With the help of these results, we conclude the proof of Proposition~\ref{prop_contibution_higher_order_terms} 
in Subsection~\ref{proof_prop_contibution_higher_order_terms}.

\subsection{Integration by parts} \label{eq-IPP_higher_order_terms}

One easily deduces from the recursion relation (\ref{eq-iterative_formula_omega_k}) that for any integer $j\geq 1$,
 \begin{eqnarray} \label{eq-exact_expression_norm_omega_k}  
  & &\| \omeg{2j+1} (t) \|^2  
 =   \\  \nonumber
 & & 
 \int_0^t \D s \int_0^t \D s ' \int_0^s \D \tau \int_0^{s'} \D \tau' \, e_{21} (s) e_{21}(s') \E^{\I  (\dynphase{1}{2}- \zeta_{12} +\theta_{12}^{-} )(s,\tau)}
 \E^{-\I  (\dynphase{1}{2} - \zeta_{12} +\theta_{12}^{-} )(s',\tau')} 
Q_j(s,\tau;s',\tau')
 \end{eqnarray}
 with
  \begin{eqnarray} \label{eq-def_Q_j}
    \nonumber
    Q_j(s,\tau; s',\tau') & \equiv & \frac{1}{e_{21}(s) e_{21}(s')} 
    \bra{\omeg{2j-1} (\tau)} \Kt{1}{2} (\tau)^\ast \Kt{2}{1} (s)^\ast  \Kt{2}{1} (s') \Kt{1}{2} (\tau')  \otimes
    \\
    & & 
    \hspace*{1cm} W (- F_{12}(s,\tau) )  W (F_{12}(s',\tau' ) ) \ket{\omeg{2j-1} (\tau')} 
    \;.
  \end{eqnarray}
  We start by deriving an exact formula for $\| \omeg{2j+1} (t) \|^2$, obtained
  through two integrations by parts in  the integrals in (\ref{eq-exact_expression_norm_omega_k}).
  
 \vspace*{2mm}

\begin{prop} \label{prop_IbP1_higher_order}
 Under assumptions~\ref{assum-gap}-\ref{assum-smoothness}, one has for any integer $j\geq 1$ and any rescaled time $t\in (0,1]$, 
  \begin{eqnarray}  \label{eq-IPP_formula_highest_order_terms}
 \| \omeg{2j+1} (t) \|^2   \nonumber
   &  =  &
 \eps^2 \re \bigg\{ \int_0^t \D \tau \int_0^t \D \tau' \, \Big( Q_j (\tau,\tau;\tau',\tau')
  - 2 \E^{- \I ( \dynphase{1}{2}-\zeta_{12}+\theta_{12}^{-} ) (t,\tau')} Q_j(\tau,\tau;t,\tau') 
        \\  \nonumber
   & &  + \E^{\I  ( \dynphase{1}{2}-\zeta_{12}+\theta_{12}^{-} ) (t,\tau)} \E^{- \I  (\dynphase{1}{2}-\zeta_{12}+\theta_{12}^{-} ) (t,\tau')}
        Q_j (t,\tau;t,\tau') \Big)
    \\  \nonumber
     & & 
    + 2  \int_0^t \D s \int_0^s \D \tau \int_0^t \D \tau '  \,\E^{\I  \dynphase{1}{2}(s,\tau)}
    \partial_s \Big( \E^{ \I (- \zeta_{12} +\theta_{12}^{-} ) (s, \tau)} \big( Q_j (s,\tau; \tau ',\tau ')
    \\  \nonumber
    & & -  \E^{- \I  ( \dynphase{1}{2}-\zeta_{12}+\theta_{12}^{-} ) (t,\tau ')} Q_j (s,\tau;t,\tau') \big) \Big)
    \\  \nonumber
    & & + \int_0^t \D s \int_0^s \D \tau \int_0^t \D s '  \int_0^{s'} \D \tau' \,\E^{\I  (\dynphase{1}{2}(s,\tau)- \dynphase{1}{2}(s',\tau'))}
    \times
    \\  
    & &
    \partial_s \partial_{s'} \Big( \E^{ \I (- \zeta_{12} +\theta_{12}^{-} ) (s, \tau)}  \E^{ -\I (- \zeta_{12} +\theta_{12}^{-} ) (s', \tau')} Q_j (s,\tau;s',\tau') \Big) 
    \bigg\} \;.
 \end{eqnarray}
\end{prop}

\vspace{2mm}

\Proof
This follows by integrating
 the two integrals over $s$ and $s'$  in (\ref{eq-exact_expression_norm_omega_k})  by parts,
 using $e_{21}(s) \E^{\I \dynphase{1}{2} (s,\tau)} = \I \eps \partial_s ( \E^{\I \dynphase{1}{2} (s,\tau)})$.
 The calculation is simplified by relying on
 \begin{equation} \label{eq-symmetry_prop_Q_j}
   Q_j(s,\tau;s',\tau') = \overline{Q_j (s',\tau'; s,\tau)}
 \end{equation}
to recognize  complex conjugate terms. 
\hfill{\finpro}

\vspace{2mm}

Combining Proposition~\ref{prop_IbP1_higher_order} with the results of Section~\ref{sec-estim_eta_zeta},
one can derive the following bound on $ \| \omeg{2j+1} (t) \|^2 $.

\vspace*{2mm}

\begin{prop} \label{prop_IbP2_higher_order}
  Let assumptions~\ref{assum-gap}-\ref{assum-decay_gamma}   hold.
Then for any integer $j\geq 1$ and $t\in (0,1]$, 
  \begin{eqnarray} \label{eq-corollary-IbP_higher_order}
  \nonumber  
  \| \omeg{2j+1} (t) \|^2
  &   \leq & 
    c_1^2 \big( \eps^2 + \lambda^2 \eps + \lambda^4 \big)
    \sup_{0\leq \tau \leq t}  \| \omeg{2j-1} (\tau)\|^2 +
    c_2^2 \big( \eps^2 + \lambda^2 \eps \big) D_{\lambda,\eps}^{(j)} (t)
    \\
    & &
    + \eps^2 \big| E_{\lambda,\eps}^{(j)} (t) \big| \;,
  \end{eqnarray}
  where the positive constants $c_1$ and $c_2$ are independent of $(\lambda, \eps, j, t)$ and we have set
  \begin{eqnarray}
    \\ \nonumber \label{eq-def_D}
    D_{\lambda,\eps}^{(j)} (t)
      & \equiv &
    \sup_{0\leq\tau'\leq s'\leq t}
     \int_0^t \D s \int_0^s \D \tau \Big(
     \Big| \partial_\sigma Q_j(s,\sigma ,\tau ; s',s',\tau')|_{\sigma=s} \Big| +
     \\ 
     & & + \Big| \partial_\sigma \partial_{\nu'} Q_j(s,\sigma,\tau ; \nu',s',\tau')|_{\sigma =s,\nu'=s'} \Big| \Big) 
    \\ \nonumber
    E_{\lambda,\eps}^{(j)} (t)
      & \equiv &
    \int_0^t \D s \int_0^s \D \tau \int_0^t \D s' \int_0^{s'} \D \tau' \, \E^{ \I (\dynphase{1}{2}- \zeta_{12} +\theta_{12}^{-} ) (s, \tau)}
    \E^{ -\I (\dynphase{1}{2} - \zeta_{12} +\theta_{12}^{-} ) (s', \tau')} \times
    \\ \label{eq-def_E}
    & &
    \partial_\sigma \partial_{\sigma'} Q_j(s,\sigma,\tau ; s',\sigma',\tau')|_{\sigma=s,\sigma'=s'} 
  \end{eqnarray}
with
  \begin{eqnarray} \label{eq-def_Q_j_bis}
    \nonumber
    Q_j(s, \sigma,\tau;s', \sigma', \tau') & \equiv & \frac{1}{e_{21}(s) e_{21}(s')} 
    \bra{\omeg{2j-1} (\tau)} \Kt{1}{2} (\tau)^\ast \Kt{2}{1} (s)^\ast  \Kt{2}{1} (s') \Kt{1}{2} (\tau')  \otimes
    \\
    & & 
    \hspace*{1cm} W (- F_{12}(\sigma ,\tau) )  W (F_{12}(\sigma ',\tau' ) ) \ket{\omeg{2j-1} (\tau')} \;.
  \end{eqnarray}
\end{prop}

\vspace{2mm}

\Proof
By Assumption~\ref{assum-smoothness} and the definition (\ref{eq-k}) of $K(s)$, the three suprema 
\begin{equation} \label{eq-def_k_infty}
  k_\infty \equiv \sup_{0 \leq s \leq 1}\| \widetilde{K}_{21} (s)\| 
  \quad ,\quad 
  k_\infty' \equiv \sup_{0 \leq s \leq 1}\| \partial_s \widetilde{K}_{21} (s)\|
  \quad \text{ and }\quad 
  \ell_\infty \equiv \sup_{0\leq s \leq 1} \big| \partial_s\ln | e_{21}|(s) \big|
\end{equation}
are finite (in fact, $\| \widetilde{K}_{21} (s)\| \leq \| {K} (s)\|$ and $\| \partial_s \widetilde{K}_{21} (s)\| \leq 2 k_\infty^2 +  \| \partial_s {K} (s)\|$ for any $s \in [0,1]$).
Let us fix  $t \in (0,1]$.
Thanks to (\ref{eq-def_Q_j}), one has
\begin{equation} \label{eq-bound_on_Q_j}
  \sup_{0 \leq \tau \leq s \leq t} \sup_{0\leq\tau'\leq s'\leq t}  | Q_j(s,\tau ; s',\tau') |
  \leq \widetilde{c}_1 \sup_{0\leq \tau \leq t}  \| \omeg{2j-1} (\tau)\|^2  
\end{equation}
with $\widetilde{c}_1=k_\infty^4/\delta^2$.
One then deduces from (\ref{eq-IPP_formula_highest_order_terms}) and (\ref{eq-symmetry_prop_Q_j})  that
\begin{eqnarray} \label{eq-proof_IbP_hiher_order}
  \nonumber
  & &  \| \omeg{2j+1} (t) \|^2 \; \leq  \;   \widetilde{c}_1 \eps^2   \big( 4 t^2 + 2 t  Z_t + Z_t^2  \big)
  \sup_{0\leq \tau \leq t}  \| \omeg{2j-1} (\tau)\|^2
    \\ 
    & &
  + \eps^2 \big(  4 t + 2  Z_t  \big) \sup_{0\leq\tau'\leq s'\leq t} \int_0^t \D s \int_0^s \D \tau \, | \partial_s Q_j(s,\tau ; s', \tau') | 
      \\ \nonumber
    & &
      + \eps^2 \bigg| \int_0^t \D s \int_0^s \D \tau \int_0^t \D s' \int_0^{s'} \D \tau'\,  \E^{ \I (\dynphase{1}{2}- \zeta_{12} +\theta_{12}^{-} ) (s, \tau)}
    \E^{ -\I (\dynphase{1}{2} - \zeta_{12} +\theta_{12}^{-} ) (s', \tau')}  \partial_s \partial_{s'} Q_j(s,\tau ; s',\tau') \bigg|
  \end{eqnarray}
with
\begin{equation*}
  Z_t \equiv   \int_0^t \D s \int_0^s \D \tau \, | \partial_s (\zeta_{12}- \theta _{12}^{-} ) (s,\tau)|\;.
\end{equation*}
In what follows, $\widetilde{c}_1$, $\widetilde{c}_2$, \ldots, denote constants independent of $(\lambda,\eps,j,t)$.
Decomposing the derivative of $Q_j$ as
\begin{equation*}
  \partial_s Q_j (s,\tau;s',\tau') =  \partial_\nu Q_j (\nu,s,\tau;s',s',\tau')|_{\nu=s} + \partial_\sigma Q_j (s,\sigma,\tau; s',s',\tau')|_{\sigma=s}
\end{equation*}
and using 
\begin{equation*}
  \Big|  \partial_\nu Q_j (\nu,s,\tau;s',s',\tau')|_{\nu=s} \Big|
  \leq  \frac{k_{\infty}^3}{\delta^2} \Big( \ell_{\infty} k_\infty + k_{\infty}'  \Big)  \sup_{0\leq \tau \leq t}  \| \omeg{2j-1} (\tau)\|^2 \;,
\end{equation*}
the supremum in the second line of
(\ref{eq-proof_IbP_hiher_order}) can be bounded from above  by
\begin{equation*} 
 \widetilde{c}_2 \sup_{0\leq \tau \leq t}  \| \omeg{2j-1} (\tau)\|^2 + 
  \sup_{0\leq\tau'\leq s'\leq t} \int_0^t \D s \int_0^s \D \tau \,\Big| \partial_\sigma Q_j(s,\sigma ,\tau ; s',s',\tau')|_{\sigma=s} \Big| \;.
\end{equation*}
Similarly, the integral in the last line of (\ref{eq-proof_IbP_hiher_order}) is bounded  by
\begin{eqnarray*}
& &  \widetilde{c}_3 \sup_{0\leq \tau \leq t}  \| \omeg{2j-1} (\tau)\|^2
  + \sup_{0\leq\tau'\leq s'\leq t}  \int_0^t \D s \int_0^s \D \tau \,\big| \partial_\sigma \partial_{\nu'} Q_j(s,\sigma,\tau ; \nu',s',\tau')|_{\sigma=s,\nu'=s'} \big| 
  \\
  & & +
  \bigg|
  \int_0^t \D s \int_0^s \D \tau \int_0^t \D s' \int_0^{s'} \D \tau' \,  \E^{ \I (\dynphase{1}{2}- \zeta_{12} +\theta_{12}^{-} ) (s, \tau)}
  \E^{ -\I (\dynphase{1}{2} - \zeta_{12} +\theta_{12}^{-} ) (s', \tau')} \times
  \\ & &
  \partial_\sigma \partial_{\sigma'} Q_j(s,\sigma,\tau ; s',\sigma',\tau')|_{\sigma=s,\sigma'=s'}
  \bigg|\;,
\end{eqnarray*}
where we have taken advantage of $Q_j(\nu,\sigma,\tau;\nu',\sigma',\tau') = \overline{Q_j(\nu',\sigma',\tau' ;\nu,\sigma ,\tau)}$.
 But by  (\ref{eq-theta12}) and (\ref{eq-def_qtrans_eta}), one has 
$\big| \partial_s \theta_{12}^{-} ( s, \tau) \big| = \big| \partial_s \theta_{21}^{+} (\tau , s) \big|
  \leq
  \big| \partial_s \eta_{21} (\tau, s ) \big|$ \footnote{The same bound holds in the positive temperature case, still with the zero temperature expression for the upper bound $|\partial_s\eta_{21}(\tau,s)|$, because the left side of the inequality (namely $\theta^\pm_{12})$ is independent of the temperature, see Section \ref{anpote}. } and thus
  \begin{equation*}
  \int_0^t \D s \int_0^s \D \tau\, \big| \partial_s \theta_{12}^{-} ( s, \tau) \big|
  \leq
   \int_0^t \D s \int_s^t \D \tau \,\big| \partial_\tau \eta_{21} (s, \tau ) \big|\;.
  \end{equation*}
Applying  Corollary~\ref{eq-corollary-bound_int_derivative_q} and   Proposition~\ref{prop-5}, this yields
  \begin{equation*}
  Z_t  
  \leq c' \frac{t^2}{2} \frac{\lambda^2}{\eps} + c \lambda^2 \eps^{m_1-1} | \ln \eps |^{\delta_{m,1}} = \Oo(\lambda^2 \eps^{-1})
  \end{equation*}
 (note that the labels $1,2$ of the energy levels can be exchanged
  without altering the results of  Corollary~\ref{eq-corollary-bound_int_derivative_q}). 
    Collecting the results above, the desired bound follows from
  (\ref{eq-proof_IbP_hiher_order}). 
  \hfill{\finpro}

\subsection{Estimation of $D_{\lambda,\eps}^{(j)}$} \label{sec-estimation_D}

\begin{prop} \label{prop_bound_on_D}
  Let assumptions~\ref{assum-smoothness}-\ref{assum-decay_gamma}   hold.
  Then for any integer $j\geq 1$ and $t\in (0,1]$, one has
\begin{equation}
\sum_{j=1}^\infty \big( D_{\lambda,\eps}^{(j)} (t) \big)^\onehalf
 \leq c \lambda \sqrt{r \Big( \frac{1}{\eps} \Big)} = \Oo\big(\lambda \eps^{\frac{m_1-1}{2}} | \ln \eps |^{\onehalf \delta_{m,1}} \big)\;.
\end{equation}
with $c>0$ independent on $(\lambda,\eps,j,t)$ and $r(1/\eps)$ defined by (\ref{eq-def_d(z)}).
\end{prop}

\vspace{2mm}

\Proof
We divide the proof into three steps.

\vspace{2mm}

\noindent {\bf STEP 1.} Let us show that for any $t \in (0,1]$,
\begin{equation} \label{eq-bound_on_D}
 \big| D_{\lambda,\eps}^{(j)} (t) \big|
 \leq c \sup_{0 \leq \tau'\leq s'\leq t} \intmult{0 \leq v_{2j-1} \leq \cdots \leq v_1 \leq \tau \leq s \leq t}
\!\!\!\!\!\! \D s \,\D \tau \, \D^{2j-1} \vv \,
  \intmult{0 \leq v_{2j-1}' \leq \cdots \leq v_1' \leq \tau'} \!\!\!\! \D^{2j-1} \vv' \,\big| \partial_s \Rrj ( s ,\tau, \vv ;s',\tau',\vv') \big|
\end{equation}
with $c$  as in the Proposition and, for any
$\vv=(v_1,\cdots,v_{2j-1}) $, $\vv' = (v_1',\cdots, v_{2j-1}' ) \in \real^{2j-1}_+$,
\begin{eqnarray} \nonumber \label{eq-def_R_j}
  \Rrj ( s,\tau, \vv ;s',\tau',\vv' )
  & \equiv  &
  \langle \chi |    
  W( -F_{12} ( v_{2j-1},v_{2j} ))\cdots W(- F_{12} ( v_{-1},v_0 ))
  \\
  & &
  W( F_{12} ( v_{-1}',v_0 ' )) \cdots W( F_{12} ( v_{2j-1}' ,v_{2j}' )) \,\chi \rangle\;,
\end{eqnarray}
where we have set $v_{2j}= 0$, $v_0 = \tau$, $v_{-1} = s$ and, similarly, $v_{2j}'= 0$,
$v_0' = \tau'$, $v_{-1}' = s'$. We shall freely pass from $(s,\tau)$ to $(v_{-1} ,v_{0})$ and so on,  wherever convenient in the sequel.

Actually, 
thanks to (\ref{eq-iterative_formula_omega_k}) and (\ref{eq-iterative_formula_omega_k_k-1}), the vector  $\omeg{2j-1} (\tau)$
  is given by the multiple integral
  \begin{eqnarray} \label{eq-Dyson_expansion_omega_2j+1}
  \nonumber  
 & &  \omeg{2j-1}(\tau)
   = 
   - \intmult{0 \leq v_{2j-1} \leq \cdots \leq v_1 \leq \tau} \D^{2j-1} \vv \, \exp \Big\{ - \I \sum_{k=1}^j ( \dynphase{1}{2}-\zeta_{12}+\theta_{12}^{-} )
   (v_{2k-1},v_{2k}) \Big\} \times
       \\
       & &  \hspace*{0mm} \Kt{2}{1} (v_1)  \Kt{1}{2} (v_2)  \cdots \Kt{2}{1} (v_{2j-1}) \otimes W( F_{12} (v_1,v_2) ) \cdots W( F_{12} ( v_{2j-1},v_{2j} ))
       \psi_1(0) \otimes \chi . 
  \end{eqnarray}
  Plugging this formula into (\ref{eq-def_Q_j_bis}) gives
\begin{eqnarray} \label{eq-Q_j_and_R_j}
  \nonumber  
 & &  Q_j ( s,\sigma,\tau; s',\sigma',\tau')  = \frac{1}{e_{21}(s) e_{21} (s')} \intmult{0 \leq v_{2j-1} \leq \cdots \leq v_1 \leq \tau} \D^{2j-1} \vv \,
  \intmult{0 \leq v_{2j-1}' \leq \cdots \leq v_1' \leq \tau'} \D^{2j-1} \vv' \,
\\ \nonumber
& & \hspace*{1cm}
\exp \Big\{ \I \sum_{k=1}^j \Big( ( \dynphase{1}{2}-\zeta_{12}+\theta_{12}^{-} ) (v_{2k-1},v_{2k}) - ( \dynphase{1}{2}-\zeta_{12}+\theta_{12}^{-} ) (v_{2k-1}',v_{2k}') \Big) \Big\} \times
\\ \nonumber
& & \hspace*{1cm}
\langle \psi_1 (0) |  \Kt{2}{1} (v_{2j-1})^\ast  \cdots \Kt{1}{2} (v_0 )^\ast \Kt{2}{1} (v_{-1})^\ast \Kt{2}{1} (v_{-1}') \Kt{1}{2} (v_0')
\cdots \Kt{2}{1} (v_{2j-1}') \, \psi_1 (0) \rangle \times
\\[2mm]
& & \hspace*{1cm} {\cal R}_{\lambda,\eps}^{(j)} ( \sigma ,\tau, \vv ;\sigma',\tau',\vv')\;.
\end{eqnarray}
The inequality (\ref{eq-bound_on_D}) then follows from Assumption~\ref{assum-smoothness} and the boundedness of $k_\infty$, $k_\infty'$, and $\ell_\infty$ in (\ref{eq-def_k_infty}).

\vspace{2mm}

\noindent {\bf STEP 2:} Exact formula for $ \Rrj (s, \tau, \vv ; s', \tau',\vv' )$.

\begin{lem} \label{prop-exact_formula_D}
One has 
\begin{eqnarray} \label{eq-exact_formula_D}
\nonumber  
& &   \Rrj (s, \tau, \vv ; s', \tau',\vv')  = 
  \E^{\I \theta(\vv,\vv')}
  \exp
  \bigg\{ - \onehalf \sum_{k=1}^j \big\langle F_{12} (v_{2k-1}, v_{2k} ) \, , \, F_{12} (s,\tau) - F_{12} (s',\tau') \big\rangle \bigg\}
  \\
&&  
\hspace*{3mm}   \times \exp \bigg\{  \onehalf \sum_{k=1}^j \big\langle F_{12} (s,\tau) - F_{12} (s',\tau')\, , \, F_{12} (v_{2k-1}', v_{2k}') \big\rangle
    + \onehalf { \big\langle F_{12} (s, \tau)\, , \, F_{12} (s',\tau') \big\rangle \bigg\}}
 \\ \nonumber
 & &
\hspace*{3mm} 
  \times \exp \bigg\{ 
      - \frac{1}{4} \bigg\| \sum_{k=1}^{j} \Big( F_{12} (v_{2k-1},v_{2k}) - F_{12} ( v_{2k-1}' , v_{2k}' ) \Big) \bigg\|^2
      - \frac{1}{4} \big\| F_{12} (s,\tau) \big\|^2 - \frac{1}{4} \| F_{12} (s',\tau') \big\|^2 \bigg\}\;,
\end{eqnarray}      
where the function $\theta(\vv, \vv'): \real^{4j-2}_+ \rightarrow \real$ is independent of $s,\tau,s'$, and $\tau'$.
\end{lem}

\vspace{2mm}

\Proofof{Lemma \ref{prop-exact_formula_D}} This follows from repeated applications of the properties (\ref{eq-product_Weyl_op}) and (\ref{eq-expectation_Weyl}) of the Weyl operators.
To get formula (\ref{eq-exact_formula_D}), one may apply the following identity, which is a consequence of these two properties:
for any $F,G,H \in L^2(\real^3)$, it holds
\begin{equation*}
\langle \chi | W (-G) W (F) W ( H) \, \chi \rangle
 = \exp
\Big\{  \onehalf \big( - \langle F \,,\,H \rangle + \langle G\,,\,F\rangle + \I \,\im   \langle G\,,\,H\rangle \big)
- \frac{1}{4} \big( \| G- H \|^2 + \| F \|^2 \big)
\Big\}\;.
\end{equation*}
Use this formula with $G=\sum_{k=1}^j F_{12}(v_{2k-1},v_{2k})$, $H=\sum_{k=1}^j F_{12}(v'_{2k-1}, v'_{2k})$ and $F=-F_{12}(s,\tau) + F_{12}(s',\tau')$. The real phase $\theta(\vv,\vv')$ comes from the phases generated by \eqref{eq-product_Weyl_op} when grouping the Weyl operators into the terms with $G$ and $H$. 
\finpro

\medskip

\noindent {\bf STEP 3:} We conclude the proof by using similar arguments as in the proof of Proposition~\ref{prop-1}.

Let 
$0=v_{2j} \leq v_{2j-1} \leq \cdots \leq v_{-1} \leq t \leq 1$ and
$0=v_{2j}' \leq v_{2j-1}' \leq \cdots \leq v_{-1}' \leq t$.
We denote by $\vvv = (v_{-1},v_0 ,\cdots, v_{2j-1}) \in \real_+^{2j+1}$, where we recall that  $ v_{-1}=s$ and $ v_0=\tau$, and use a similar notation with the primes.
Recalling that $F_{12} (s,\tau)= \int_\tau^s \D x\, f_{12} (x)$, one finds 
thanks to Lemma~\ref{prop-exact_formula_D} that 
\begin{eqnarray*}
\partial_s \Rrj (\vvv, \vvv' )   
&  = &
\onehalf
  \bigg( - \sum_{k=1}^j \big\langle F_{12} (v_{2k-1}, v_{2k} ) \,,\,f_{12}(s) \big\rangle
    - \re \big\langle F_{12}(s,\tau)\,,\,f_{12}(s)\big\rangle
  \\
  & & \hspace*{2mm}
  +  \sum_{k=0}^j  \big\langle f_{12}(s)  \,,\, F_{12} (v_{2k-1}' , v_{2k}' )\big\rangle  \bigg) \Rrj ( \vvv , \vvv' ) \;.
\end{eqnarray*}
(Note that the second sum starts with $k=0$.)
Using  $| \Rrj (\vvv, \vvv' )  | \leq 1$ (see (\ref{eq-def_R_j})), 
\begin{eqnarray*}
2 \big| \partial_s \Rrj ( \vvv, \vvv' ) \big|    
& \leq  &
 \sum_{k=1}^j \big| \big\langle F_{12} (v_{2k-1}, v_{2k} ) \,,\,f_{12}(s) \big\rangle \big|
 + \big| \re \big\langle F_{12} (s, \tau ) \,,\,f_{12}(s) \big\rangle \big|
\\
& &
+  \sum_{k=0}^j \big| \big\langle  F_{12} (v_{2k-1}' , v_{2k}' )  \,,\, f_{12}(s)  \big\rangle \big|
\;.
\end{eqnarray*}

Arguing as in the proof of Proposition~\ref{prop-1}, one finds 
\begin{eqnarray} \label{eq-proof_prop5.3}
\nonumber  
& & \big| \partial_s \Rrj (\vvv, \vvv' ) \big|    
\leq 
\frac{2 M^2 \lambda^2}{\eps}
\bigg\{ \sum_{k=1}^j  \bigg| \int_{\frac{v_{2k}-s}{\eps}}^{\frac{v_{2k-1}-s}{\eps}} \D x \,\gamma(x) \bigg| 
 +  \bigg| \int^{0}_{\frac{\tau-s}{\eps}} \D x \,\gamma_R (x) \bigg| 
  \\
  & &
  \hspace*{5mm}
  + \sum_{k=0}^j  \bigg| \int_{\frac{v_{2k}' -s}{\eps}}^{\frac{v_{2k-1}' -s}{\eps}} \D x \,\gamma(x) \bigg| 
   \bigg\}
  + 4 M^2 \lambda^2
  \bigg\{
  \int_{-\frac{s}{\eps}}^{0} \D x \,|x \gamma (x)|   + \int_{-\frac{s}{\eps}}^{\frac{s'-s}{\eps}} \D x \,|x\gamma (x)|
  \bigg\}\;.
\end{eqnarray}
One has
\begin{equation*}
  \sum_{k=1}^j  \left| \int_{\frac{v_{2k}-s}{\eps}}^{\frac{v_{2k-1}-s}{\eps}} \D x \,\gamma(x) \right|
  \leq  \Gamma \bigg( \frac{s-\tau}{\eps} \bigg) 
\end{equation*}
and, for any $k=0,\ldots,j$,
\begin{equation*}
  \left| \int_{\frac{v_{2k}' -s}{\eps}}^{\frac{v_{2k-1}' -s}{\eps}} \D x \,\gamma(x) \right|
  \leq \Gamma \bigg( \frac{|v_{2k}'-s|}{\eps} \bigg) +  \Gamma \bigg( \frac{|v_{2k-1}' -s|}{\eps} \bigg)\;,
\end{equation*}
where we have set
\begin{equation}
  \Gamma ( y ) \equiv \int_y^\infty \D x \, | \gamma (x)| = \int_{-\infty}^{-y} \D x\,| \gamma(x) |
\end{equation}
and the second bound follows from  $\int_{-\infty}^\infty \D x \,\gamma (x) = 0$ see (\ref{eq-vanishing_int_gamma_R}). 
One shows with the help of the change of variables $y = |v' - s|/\eps$ that for any $0 \leq v' \leq t$,
\begin{equation*}
  \int_0^t \D s \, \Gamma \bigg( \frac{|v'-s|}{\eps} \bigg)  \leq 2 \eps \int_0^{\frac{t}{\eps}} \D y \,\Gamma (y)\;.
\end{equation*}
Thanks to  (\ref{eq-proof_prop5.3}), the three last bounds, and $\int_{0}^\infty \D x\,\gamma_R (x)=0$, one is led to
\begin{eqnarray*}
\nonumber  
& & \sup_{0 \leq \tau'\leq s'\leq t} \intmult{0 \leq v_{2j-1} \leq \cdots \leq v_{-1} \leq t} \D^{2j+1} \vvv \intmult{0 \leq v_{2j-1}' \leq \cdots \leq v_1' \leq t} \D^{2j-1} \vv'   \,
\big| \partial_s \Rrj ( \vvv , \vvv' ) \big|
\\
& & \hspace*{5mm}
\leq \frac{2 M^2 \lambda^2}{ ((2j-1)!)^2} \bigg( ( 6 + 4 j) \int_0^{\frac{1}{\eps}} \D y \, \Gamma (y) + 6 \int_0^{\frac{1}{\eps}} \D x x |\gamma(x)|
\bigg)
\leq c_j \lambda^2 r \Big( \frac{1}{\eps} \Big)\;,
\end{eqnarray*}
where $c_j>0$ depends on $j$ only and satisfies $\sum_{j\geq 1} \sqrt{c_j} < \infty$.
  Substituting this bound into (\ref{eq-bound_on_D}) and relying on (\ref{eq-bound_on_r}), one gets the result of
Proposition~\ref{prop_bound_on_D}.
  \finpro

\subsection{Estimation of $E_{\lambda,\eps}^{(j)}$} \label{sec-estimation_E}

Replacing (\ref{eq-Dyson_expansion_omega_2j+1}), (\ref{eq-def_Q_j_bis}), and (\ref{eq-def_R_j}) into (\ref{eq-def_E}), it follows
\begin{eqnarray} \label{eq_expression_E}
\nonumber  
& &   E_{\lambda,\eps}^{(j)} (t)  =  \intmult{0 \leq v_{2j-1} \leq \cdots \leq v_{-1} \leq t} \D^{2j+1} \vvv  \intmult{0 \leq v_{2j-1}' \leq \cdots \leq v_{-1}' \leq t} \D^{2j+1} \vvv '\,
  \frac{1}{e_{21}(v_{-1}) e_{21}(v_{-1}')} \times
  \\ 
  & & \hspace*{1cm}
  \exp \bigg\{ \I \sum_{k=0}^j \big( ( \dynphase{1}{2} - \zeta_{12} + \theta_{12}^{-} ) (v_{2k-1},v_{2k} )
  - ( \dynphase{1}{2} - \zeta_{12} + \theta_{12}^{-} ) (v_{2k-1}' , v_{2k}' ) \big) \bigg\} \times
  \\[2mm] \nonumber
  & &  \hspace*{1cm}
  \langle \psi_1 (0) | \widetilde{K}_{21} ( v_{2j-1})^\ast \cdots \widetilde{K}_{21} (v_{-1}) ^\ast
  \widetilde{K}_{21} ( v_{-1}' ) \cdots \widetilde{K}_{21} (v_{2j-1}' ) \, \psi_1 (0) \rangle  \partial_{v_{-1}} \partial_{v_{-1}'}  \Rrj ( \vvv , \vvv' )\;.
\end{eqnarray}
Now, according to Lemma~\ref{prop-exact_formula_D} (recall that $s = v_{-1}$, $\tau=v_{0}$, $s'=v_{-1}'$, and $\tau'=v_0'$), one has
\begin{eqnarray} \label{eq-second_derivative_R}  
  \nonumber
  & & \partial_s\partial_{s'} \Rrj ( \vvv , \vvv' )  
  = \frac{1}{4}
  \bigg\{
  \bigg[ - \sum_{k=1}^j \big\langle F_{12} (v_{2k-1}, v_{2k} ) \,,\,f_{12}(s) \big\rangle
    - \re \big\langle F_{12}(s,\tau)\,,\,f_{12}(s)\big\rangle
\\ \nonumber
& & \hspace*{5mm}
  +  \sum_{k=0}^j  \big\langle f_{12}(s)  \,,\, F_{12} (v_{2k-1}' , v_{2k}' )\big\rangle  \bigg]
  \bigg[  \sum_{k=0}^j \big\langle F_{12} (v_{2k-1}, v_{2k} ) \,,\,f_{12}(s') \big\rangle
    - \re \big\langle F_{12}(s',\tau')\,,\,f_{12}(s')\big\rangle
\\
& &  \hspace*{5mm}
- \sum_{k=1}^j  \big\langle f_{12}(s')  \,,\, F_{12} (v_{2k-1}' , v_{2k}' )\big\rangle  \bigg]
  + 2 \big\langle f_{12} (s')\,,\,f_{12}(s) \big\rangle
\bigg\}  \Rrj (\vvv , \vvv' ) \;.
\end{eqnarray}
The term involving the scalar product $\langle f_{12} (s')\,,\,f_{12}(s) \rangle$ requires some special care.
Its contribution to $ E_{\lambda,\eps}^{(j)} (t)$ is given by
\begin{eqnarray*}
  G_{\lambda,\eps}^{(j)} (t)
& \equiv & \onehalf \int_0^t \D s \int_0^s \D \tau \int_0^t \D s' \int_0^{s'} \D \tau' 
  \E^{\I ( \dynphase{1}{2} - \zeta_{12} + \theta_{12}^{-} ) (s,\tau )}  \E^{-\I ( \dynphase{1}{2} - \zeta_{12} + \theta_{12}^{-} ) (s',\tau' )}
  \big\langle f_{12} (s')\,,\,f_{12}(s) \big\rangle \times
  \\
& & Q_{j} (s,\tau; s',\tau')
\end{eqnarray*}
and in view of (\ref{eq-bound_on_Q_j}) can be bounded for any $t \in (0,1]$ as follows 
\begin{equation} \label{eq-bound_on_G}
  | G_{\lambda,\eps}^{(j)} (t) |
   \leq  \onehalf
  \widetilde{c}_1 \sup_{0 \leq \tau \leq t} \| \omeg{2j-1} (\tau) \|^2 \int_0^{1} \D s \int_0^{1} \D s'
  \big| \big\langle f_{12} (s')\,,\,f_{12}(s) \big\rangle \big|
   \leq 
  c_3^2 \frac{\lambda^2}{\eps} \sup_{0 \leq \tau \leq t} \| \omeg{2j-1} (\tau) \|^2
\end{equation}  %
with $c_3^2 = 4 M^2 \widetilde{c}_1 \int_0^\infty \D x\,|\gamma| (x) < \infty$. 
The contribution to $E_{\lambda,\eps}^{(j)}(t)$ of the other terms in the derivative (\ref{eq-second_derivative_R})  is controlled
in the following Proposition.

\begin{prop} \label{prop_bound_on_F}
  Let assumptions~\ref{assum-gap}-\ref{assum-decay_gamma}   hold and let us set 
  $ \widetilde{E}_{\lambda,\eps}^{(j)} (t) = E_{\lambda,\eps}^{(j)} (t) - G_{\lambda,\eps}^{(j)} (t)$.
  Then for any integer $j\geq 1$ and $t\in (0,1]$, one has
\begin{equation}
\sum_{j=1}^\infty \big| \widetilde{E}_{\lambda,\eps}^{(j)} (t) \big|^\onehalf
 = \Oo \big(\lambda^2 \eps^{\min \{ m-1\,,\,-\onehalf\} } | \ln \eps |^{\onehalf \delta_{m,1/2}} \big)\;.
\end{equation}
\end{prop}

\vspace{2mm}

\Proof
Thanks to (\ref{eq_expression_E}) and (\ref{eq-second_derivative_R}), one has
\begin{eqnarray}
  \nonumber
& &   \big| \widetilde{E}_{\lambda,\eps}^{(j)} (t) \big|
 \leq 
  \frac{k_\infty^{4j+2}}{4 \delta^2}  \intmult{0 \leq v_{2j-1} \leq \cdots \leq v_{-1} \leq t} \D^{2j+1} \vvv  \intmult{0 \leq v_{2j-1}' \leq \cdots \leq v_{-1}' \leq t} \D^{2j+1} \vvv ' \,
    \\ \nonumber
  & &
 \bigg[ \sum_{k=1}^j \big| \big\langle F_{12} (v_{2k-1}, v_{2k} ) \,,\,f_{12}(s) \big\rangle \big|   + \big| \re \big\langle F_{12}(s,\tau)\,,\,f_{12}(s)\big\rangle \big|
   +  \sum_{k=0}^j  \big| \big\langle F_{12} (v_{2k-1}' , v_{2k}' )    \,,\, f_{12}(s)\big\rangle \big| \bigg] \times
 \\ \nonumber
 & &
 \bigg[  \sum_{k=0}^j \big| \big\langle F_{12} (v_{2k-1}, v_{2k} ) \,,\,f_{12}(s') \big\rangle \big|
    + \big| \re \big\langle F_{12}(s',\tau')\,,\,f_{12}(s')\big\rangle \big| + \sum_{k=1}^j  \big| \big\langle  F_{12} (v_{2k-1}', v_{2k}' )  \,,\, f_{12}(s') \big\rangle \big|
    \bigg]\;.
\end{eqnarray}
Proceeding as in Step 3 of the proof of Proposition~\ref{prop_bound_on_D}, one is led to
\begin{eqnarray*}
  \nonumber
\big| \widetilde{E}_{\lambda,\eps}^{(j)} (t) \big|
& \leq &  
 c k_\infty^{4j+2}\, \frac{\lambda^4}{\eps^2} \intmult{0 \leq v_{2j-1} \leq \cdots v_{-1}  \leq t} \D^{2j+1} \vvv
 \intmult{0 \leq v_{2j-1}' \leq \cdots \leq v_{-1}' \leq t} \D^{2j+1} \vvv' \,  \bigg[ 2 \Gamma \bigg( \frac{v_{-1}-v_0}{\eps} \bigg) +
    \\ \nonumber
  & &
 + \sum_{l=-1}^{2j}  \Gamma \bigg( \frac{|v_{-1}-v_l'|}{\eps} \bigg)   +     6 \eps \int_0^{\frac{t}{\eps}} \D x \,x|\gamma(x) | \bigg]
 \bigg[ 2 \Gamma \bigg( \frac{v_{-1}'-v_0'}{\eps} \bigg)  + 
   \\
 & &   
   + \sum_{l=-1}^{2j}  \Gamma \bigg( \frac{ |v_{-1}'-v_l|}{\eps} \bigg)  +   6 \eps \int_0^{\frac{t}{\eps}} \D x \,x|\gamma(x) | \bigg]
\end{eqnarray*}
yielding, for any $t \in (0,1]$,
\begin{equation*}
 \big| \widetilde{E}_{\lambda,\eps}^{(j)} (t) \big|    
      \leq
      \frac{k_\infty^{4j+2} \lambda^4}{((2j-1)!)^2} \bigg\{ c_j \Big[ r \Big( \frac{1}{\eps} \Big) \Big]^2 + \frac{2 c}{\eps} \int_0^{\frac{1}{\eps}}\D y \, \Gamma (y)^2 \bigg\} 
\;,      
 \end{equation*}
where the constant $c_j$ is quadratic in $j$.
Using Assumption~\ref{assum-decay_gamma}, one easily shows that
\begin{equation}
 \int_0^{\frac{1}{\eps}}\D y \, \Gamma(y)^2   
 =
 \begin{cases}
   \Oo  ( 1) & \text{ if $m>\onehalf$} \\
   \Oo  ( | \ln \eps | ) & \text{ if $m=\onehalf$}\\
   \Oo  ( \eps^{2m-1} ) & \text{ if $m < \onehalf$.}
 \end{cases}  
 \end{equation}
 The result follows.
 \finpro

\subsection{End of the proof of Proposition~\ref{prop_contibution_higher_order_terms}}
\label{proof_prop_contibution_higher_order_terms}

Combining the results of Propositions~\ref{prop_IbP2_higher_order}, \ref{prop_bound_on_D}, and~\ref{prop_bound_on_F}
and taking advantage of (\ref{eq-bound_on_G}), one gets 
\begin{eqnarray} \label{eq-proof_prop5.1}
\nonumber  
 & &  \sum_{j=1}^\infty \sup_{0 \leq t \leq 1} \big\| \omeg{2j+1} (t) \big\|
 \leq 
  \big( c_1 \eps + (c_1+c_3) \lambda \sqrt{\eps} + c_1 \lambda^2 \big)
  \sum_{j=1}^\infty \sup_{0 \leq \tau \leq 1} \big\| \omeg{2j-1} (\tau) \big\|
  \\
  & & \hspace*{1cm}
  + c_2' \lambda \eps^{\frac{m_1+1}{2}} |\ln \eps |^{\onehalf \delta_{m,1}} + c_2' \lambda^2 \eps^{\onehalf m_1} | \ln \eps |^{\onehalf \delta_{m,1}} 
   + c_4 \lambda^2 \eps^{\min \{ m,\onehalf\}} | \ln \eps |^{\onehalf \delta_{m,{1/2}}}\;.
\end{eqnarray}   
Since $\onehalf m_1 \leq \min \{ m, \onehalf\}$, the last term is much smaller  than the previous one. 
Decomposing the infinite series in the \RHS of (\ref{eq-proof_prop5.1}) as its first term plus the remainder and noting that the latter coincides with the series in the \LHS
this gives 
\begin{eqnarray*}
  \sum_{j=1}^\infty \sup_{0 \leq t \leq 1} \big\| \omeg{2j+1} (t) \big\|
 & = &
  \Big( 1 + \Oo(\eps+ \lambda \sqrt{\eps} +\lambda^2) \Big)
  \Big( \sup_{0 \leq t \leq 1} \big\| \omeg{1} (t ) \big\|  \Oo(\eps+ \lambda \sqrt{\eps} +\lambda^2) + 
\\  
& &
\Oo\big( \lambda \eps^{\frac{m_1+1}{2}}  |\ln \eps |^{\onehalf \delta_{m,1}}  + \lambda^2 \eps^{\onehalf m_1}  |\ln \eps |^{\onehalf \delta_{m,1}}   \big)
\Big)\;.
\end{eqnarray*}
But  
$ 
\sup_{0 \leq t\leq 1}\| \omeg{1} (t) \|= \Oo(\eps+\lambda \eps^{\onehalf } +\lambda^2\eps^{\onehalf m_1} | \ln \eps |^{\onehalf \delta_{m,1}} +\lambda^3 )
$ 
by Proposition~\ref{prop-6}.

Noting furthermore that $\lambda \eps^{\frac{3}{2}}$ is much smaller than $\lambda \eps^{\frac{m_1+1}{2}} |\ln \eps |^{\onehalf \delta_{m,1}} $ and that
$\lambda^2 \eps$, $\lambda^3 \sqrt{\eps}$ and $\lambda^4 \eps^{\onehalf m_1} | \ln \eps |^{\onehalf \delta_{m,1}}$ are much smaller than $\lambda^2 \eps^{\onehalf m_1} | \ln \eps |^{\onehalf \delta_{m,1}}$, this yields 
the result of Proposition~\ref{prop_contibution_higher_order_terms}.
\finpro


\section{Positive temperatures} \label{postemp}

\subsection{Positive temperature representation} To describe the reservoir state $\omega_{R,\beta}$ at positive temperature $T=1/\beta>0$ one takes the thermodynamic limit of finite volume Gibbs states of the free bose gas, see \cite{Bratteli} or \cite{M} for example. In this limit, the expectation of a Weyl operator $W(f)$, $f\in L^2({\mathbb R}^3)$, is calculated to be 
\begin{equation}
	\label{m4}
	\omega_{R,\beta}(W(f)) = \exp \left\{ -\tfrac14 \int_{{\mathbb R}^3} |f(k)|^2 \coth(\beta \omega(k)/2)\, \D^3 k \right\}\;.
\end{equation}
This reduces to the value \eqref{eq-expectation_Weyl} for $\beta\rightarrow \infty$. For simplicity, we restrict attention to $\omega(k)=|k|$ in this positive temperature section. Following \cite{AW, JaPi}, a Hilbert space supporting the state $\omega_{R,\beta}$ as a rank-one density matrix $|\chi\rangle\langle \chi|$ is given by ${\mathcal F}_+(L^2({\mathbb R}\times S^2))$, the Fock space over the (new) single particle Hilbert space $L^2({\mathbb R}\times S^2)$, with $\chi$ denoting its vacuum vector. More precisely, we have 
\begin{equation}
\label{m7}
\omega_{R,\beta}(W(f)) = \langle \chi | W(f_\beta) \chi\rangle,
\end{equation}
where the function $f_\beta\in L^2({\mathbb R}\times S^2)$ is constructed from $f\in L^2({\mathbb R}^3)$ by the rule
\begin{equation}
	\label{m5}
	f_\beta(u,\sigma) = \sqrt{\frac{u}{1-e^{-\beta u}}}\,  |u|^{1/2} \left\{
	\begin{array}{ll}
		f(u,\sigma) & u\ge 0\\
		- \bar f(-u,\sigma) & u<0
	\end{array}
	\right..
\end{equation}
The function $f$ on the right side in \eqref{m5} is represented in polar coordinates ${\mathbb R^3}\ni k\mapsto (u,\sigma)\in [0,\infty)\times S^2$. In particular, $u=|k|$ for $u\ge 0$. The radial argument of the function $f_\beta$ on the left side is $u\in\mathbb R$.

The operator $W(f_\beta)$ in \eqref{m7} is the {\it represented  Weyl operator} acting on ${\mathcal F}_+(L^2({\mathbb R}\times S^2))$, given by 
\begin{equation}
\label{m8}
W(f_\beta) = e^{i \phi(f_\beta)},\qquad \phi(f_\beta) = \frac{1}{\sqrt 2} \big( a^*(f_\beta) +a(f_\beta) \big).
\end{equation}
Here, $a^*(f_\beta)$ and $a(f_\beta)$ are the creation and annihilation operators acting on ${\mathcal F}_+(L^2({\mathbb R}\times S^2))$, satisfying the canonical commutation relations $[a(f_\beta), a^*(g_\beta)] = \langle f_\beta,g_\beta\rangle_{L^2({\mathbb R}\times S^2)} $. 

We may write $W(f_\beta)=\pi_\beta(W(f))$, where $\pi_\beta$ is a $*$-representation of the Weyl algebra. In particular, due to  \eqref{eq-product_Weyl_op},
$$
W(f_\beta) W(g_\beta) = \pi_\beta\big(W(f)W(g)\big) = e^{-\frac i2 {\rm Im} \langle f,g\rangle}\pi_\beta\big( W(f+g)\big) = e^{-\frac i2 {\rm Im} \langle f,g\rangle} W(f_\beta+g_\beta).
$$
On the other hand, the left hand side equals $e^{-\frac i2 \langle f_\beta,g_\beta\rangle} W(f_\beta+g_\beta)$ and it is indeed easy to see directly from the definition \eqref{m5} that 
\begin{equation}
\label{m9}
{\rm Im }\langle f_\beta,g_\beta\rangle = {\rm Im} \langle f, g\rangle. 
\end{equation}
(The two inner products are in different spaces but it is clear which ones they are.) 

We assume here that the radial function $u\mapsto \omega(u)$, originally defined for $u\ge 0$ (namely, $u=|k|$) extends to $u\in\mathbb R$ so that $\omega(-u) = -\omega(u)$, the typical example being $\omega(u)=u$. Then it is readily seen from \eqref{m5} that the dynamics $t\mapsto W(e^{i \omega t}f)$ is implemented as
\begin{equation}
	\label{m10}
t\mapsto W\big((e^{i\omega t}f)_\beta\big) = e^{i t L_R}  W(f_\beta) e^{-i tL_R},
\end{equation}
where the Liouville operator $L_R$ is the second quantization of the operator of multiplication by $\omega(u)$, which can be written as (compare to \eqref{m12})
\begin{equation}
\label{m11}
L_R =\int_{{\mathbb R}\times S^2} \omega(u) a^*(u,\sigma) a(u,\sigma) \, \D u \,\D^2 \sigma.
\end{equation}
Here, $\D^2 \sigma$ is the uniform measure on $S^2$. The operator $L_R$ is the generator implementing the Bogoliubov transformation $a^*(f_\beta)\mapsto a^*(e^{i\omega t}f_\beta)$. 
\medskip

\subsection{Positive temperature setup}
\label{ptfsect}

According to the previous section, the setup for the positive temperature case is obtained from the zero temperature situation by making the following replacements.

\begin{itemize}
\item[$\bullet$] The Hilbert space \eqref{m1} is replaced by
\begin{equation}
	\label{m2}
	{\mathcal H}_{\rm tot} = {\mathbb C}^2\otimes 
	{\mathcal F}_+(L^2({\mathbb R}\times S^2)).
\end{equation}
\item[$\bullet$] The Hamiltonian $H_R$, \eqref{m12}, is replaced by the Liouvillian $L_R$, \eqref{m11}.

\item[$\bullet$] The interaction \eqref{H_int} is replaced by the operator 
\begin{equation} \label{H_int'}
	\Hint (\eps \tp ) = \lambda B ( \eps \tp ) \otimes \phi (g_\beta),
\end{equation}
where $\phi (g_\beta)$ acts on ${\mathcal F}_+(L^2({\mathbb R}\times S^2))$.

\item[$\bullet$] The initial state \eqref{eq-init_state} is replaced by
\begin{equation}
	\label{m3}
	\rho(0) = |\psi_1(0)\rangle\langle \psi_1(0)| \otimes |\chi\rangle\langle\chi|,
\end{equation}
where $\chi$ is the vacuum vector in ${\mathcal F}_+(L^2({\mathbb R}\times S^2))$. 
\end{itemize}

None of the quantities involving the two-level system only are changed (such as $H_S(t)$, $B(t)$, $K(t)$, $W_K(t)\ldots$). The transition probability  $p_{1\rightarrow 2}^{(\lambda,\varepsilon)}(t)$ is still given by the formula 
\eqref{eq-transition_proba}, where the trace is that of the space \eqref{m2}, and in which $U_{\lambda,\varepsilon}(t)$ still obeys equation \eqref{eq-Schrodinger_Ueps}, simply with $\phi(g)$ replaced by $\phi(g_\beta)$ and $H_R$ by $L_R$, see also \eqref{eq-transition_proba-postemp}. 
\medskip

The reservoir autocorrelation function \eqref{eq-gamma_def} now reads 
\begin{eqnarray} 
\label{m13}
\gamma^{\beta}(t)&=& 2\,  \omega_{R,\beta}\big( \phi(\E^{\I\omega t}g)\phi(g) \big) = 2 \big\langle\chi | \phi\big(\E^{\I u t}g_\beta\big)\phi(g_\beta)\chi \big\rangle = \big\langle\chi |  a\big(\E^{\I u t}g_\beta\big) a^*(g_\beta) \chi\big\rangle=\scalprod{\E^{\I u t} g_\beta}{g_\beta}\nonumber\\
&=& \int_0^\infty \int_{S^2}\  \frac{u^2}{\E^{\beta u}-1}\ |g(u,\sigma)|^2 \big( \E^{-\I ut} \E^{\beta u} + \E^{\I ut}\big) \,\D u \, \D^2 \sigma \; .
\end{eqnarray} 
To obtain the last equality we used \eqref{m5}. Taking the real and imaginary parts,
\begin{eqnarray}
\gamma^{\beta}(t) &=& \gamma_R^\beta(t) +\I \gamma_I^\beta (t) \nonumber\\
\gamma^{\beta}_R(t) &=& {\rm Re}\, \int_0^\infty \E^{-\I \omega t} \omega^2 \coth(\beta\omega/2) \int_{S^2}|g(\omega,\sigma)|^2 \,\D \omega \,\D^2 \sigma
\label{m14}\\
\gamma^{\beta}_I(t) &=& {\rm Im}\, \int_0^\infty \E^{-\I \omega t} \omega^2 \int_{S^2}|g(\omega,\sigma)|^2 \,\D\omega \,\D^2 \sigma \;. \label{m15}
\end{eqnarray}
The real part depends on $\beta$ but the imaginary part does not and is the same as for zero temperature. Compare with \eqref{eq-gamma_def}, \eqref{2.17}. 
 
 \medskip

\subsection{Proof of Theorem \ref{thm-1} (i)}  \label{anpote}

The analysis of Sections \ref{sec-exact_calculations},  \ref{sec-contribution_first_term}  and \ref{sec-higher_order_terms} carries through in the positive temperature case, upon making the changes \eqref{m2}-\eqref{m3}. This is so because the contribution of the reservoir is dealt with entirely in a representation independent way. For instance, the crucial result of Lemma \ref{lem-1} still holds. Indeed, \eqref{eq-Schrodinger_eq_for_Psi}  is valid with $\phi(g)$ replaced by $\phi(g_\beta)$ and $H_R$ replaced with $L_R$. The same holds for \eqref{m17}. To solve equation \eqref{m17} we use again the commutation relation \eqref{eq-Wphi} which holds for $F,G\in L^2({\mathbb R}\times S^2)$  and the ensuing relation \eqref{eq-derweyl}, where now $F(t)$, $\zeta_j(t)$ and $f_j(t)$ are given by \eqref{eq-zetaj_Fj}-\eqref{eq-fj} but with $g$ replaced by $g_\beta$. Explicitly, for example,
$f_j(t)$ becomes
$$
[f_j(t)]_\beta(u) = -\frac{\lambda}{\varepsilon} b_j(t) \E^{\frac{\I  u t}{\varepsilon} }g_\beta(u,\sigma) \in L^2({\mathbb R}\times S^2)\;
$$
Incidentally,  $ \zeta_j(t)$, \eqref{eq-zetaj_Fj}, is independent of $\beta$, as follows from \eqref{m9}. In the same vein, $\theta_{12}^\pm (s,\tau)$ defined in \eqref{eq-theta12} is independent of $\beta$ (and takes the same value as the zero temperature case).

The main term, $\| \omeg{1} (t) \|^2 $, is then given in \eqref{eq-exact_formula_norm_omega_1} and the only difference with the zero temperature case is that the real part of  $\eta_{12}$ now depends on $\beta$. The expression \eqref{eq-exact_expression_eta} of $\eta_{12}$ is still valid but now $\gamma_R(t)$ is  replaced by $\gamma^\beta_R(t)$, \eqref{m14}, (while $\gamma_I(t)$ is replaced by $\gamma^\beta_I(t)$, \eqref{m15}, and is the same as for zero temperature).

In terms of the properties of the reservoir, the analysis in Sections \ref{sec-contribution_first_term} and \ref{sec-higher_order_terms} relies entirely on assumption \ref{assum-decay_gamma}  (other than the properties $\gamma_R(-t)=\gamma_R(t)$ and $\gamma_I(-t) =-\gamma_I(t)$ which are satisfied for \eqref{m14}, \eqref{m15}). So we should now verify that \ref{assum-decay_gamma} holds  for non trivial form factors, {\it i.e.}, that  
\begin{equation}
	\label{m18}
\dss \sup_{t \in \real} ( 1 + t^2)^{\frac{m+1}{2}} |\gamma^\beta(t)|  < \infty \mbox{\quad and\quad} 
\dss \lim_{\omega \to 0+} \frac{\widehat{\gamma}^\beta ( \omega)}{\omega^m} \equiv \gamma_0\geq 0
\end{equation}
for some $m>0$. We consider again a radially symmetric $g$ of the form (see \eqref{eq-form_factor})  
\begin{equation} 
\label{m19}
g(k) = g_0 |k|^{\frac{\mu}{2}-1} \exp \Big( - \frac{|k|}{2  \omega_D} \Big)
\end{equation}
for some $\mu>0$.

  To show that the first condition in (\ref{m18}) is satisfied, we let
$\ell\in\mathbb N$, use $\E^{-\I\omega t} = \frac{1}{(-\I t)^\ell} \partial_\omega^\ell \E^{-\I\omega t}$ and integrate by parts $\ell$ times to get that 
\begin{equation}
\label{m20}
\big| \gamma_R^\beta (t) \big| \leq
\bigg| \int_0^\infty \E^{-\I \omega t} \omega^2 \coth(\beta\omega/2) \int_{S^2}|g(\omega,\sigma)|^2 \,\D \omega \,\D^2\sigma\bigg| \leq  \frac{C}{(1+t^2)^{\ell/2}}\;,
\end{equation}
provided $\mu > \ell$. More precisely, the boundary terms all vanish,
\begin{equation}
\label{m21}
\partial_\omega^r \Big(\omega^\mu \coth(\beta\omega/2) \E^{-\omega/\omega_D} \Big)\Big|_0^\infty =0,\qquad r=0,1,\ldots,\ell-1,
\end{equation}
and the final integral left over after the integrations by part is absolutely convergent. Note that $\coth(\beta\omega/2)$ has a $1/\omega$ singularity at the origin and is bounded for large $\omega$. The same argument holds to bound $\big| \gamma_I^\beta (t) \big|$, replacing the cotangent by $1$ in the integral in \eqref{m20} (in fact then, we only need $\mu > \ell-1$ since the singularity of the cotangent is absent -- we may also use the explicit formula \eqref{eq-exact_expression_gamma} in this case). We conclude that by choosing $\mu > m+1$ in \eqref{m19}, the first condition in \eqref{m18} is satisfied.

Next we turn to the second condition in (\ref{m18}).
We note that
$\widehat\gamma^\beta(\omega)= \widehat \gamma^\beta_R(\omega) +\I \widehat \gamma^\beta_I(\omega)$ with $\widehat \gamma^\beta_R$ and
$\widehat \gamma^\beta_I$ the Fourier transforms of $\gamma^\beta_R$ and $\gamma^\beta_I$, respectively. Now, by (\ref{m14}),
\begin{equation}
\label{m22}
\widehat \gamma^\beta_R(\omega) = \int_{\mathbb R} \E^{\I \omega t} \gamma^\beta_R(t) \,\D t
= \frac{1}{4\pi}\int_{\mathbb R} \E^{-\I \omega t} \int_0^\infty \big( \E^{-\I u t}+\E^{\I u t}\big) \coth(\beta u /2)\widehat \gamma(u) \,\D u \, \D t,
\end{equation}
where we recall that $\widehat \gamma(\omega)= 2\pi \omega^2 \int_{S^2}|g(\omega,\sigma)|^2 \D^2 \sigma$ for $\omega \geq 0$.
Using the representation $\int_{\mathbb R}\E^{\I \xi t} \D t =2\pi \delta(\xi)$ of the Dirac distribution we obtain from \eqref{m22}
\begin{equation}
\label{m23}
\widehat \gamma^\beta_R(\omega)= \onehalf \coth(\beta|\omega|/2)\, \widehat \gamma(|\omega|)\;,\quad \omega\in\mathbb R.
\end{equation}
We proceed in the same way to find 
\begin{equation}
\label{m24}
\widehat \gamma^\beta_I(\omega)= - \frac{\I}{2} \, {\rm sgn}(\omega) \, \widehat \gamma(|\omega|)\; ,\quad \omega\in\mathbb R\; ,
\end{equation}
where ${\rm sgn}(\omega) = 1$ if $\omega> 0$, ${\rm sgn}(\omega)=0$ if $\omega=0$, and ${\rm sgn}(\omega) = -1$ if $\omega<0$.
To satisfy the second condition in \eqref{m18} we thus require $\widehat \gamma(\omega)/\omega^{m+1}$ to have a finite limit as $\omega\rightarrow 0+$. In terms of \eqref{m19}, it suffices to take $\mu\ge m+1$.

We conclude that Assumption \ref{assum-decay_gamma} is satisfied in the positive temperature case for form factors \eqref{m19} with $\mu> m+1>1$.


\end{document}